\documentclass[11pt,a4paper]{article}
\usepackage[utf8]{inputenc}
\usepackage{graphicx,amsmath,amsthm,latexsym,amssymb}
\usepackage{float,epsfig,multirow,rotating,times}
\usepackage{multirow,subfig}
\usepackage{upgreek,wrapfig}
\usepackage{comment}
\usepackage{mathrsfs}
\usepackage{bm}
\usepackage{algorithm}
\usepackage[noend]{algpseudocode}
\usepackage{enumitem}
\usepackage[left=1in, right=1in, top=1in, bottom=1in]{geometry}
\usepackage[colorlinks,citecolor=blue,urlcolor=blue,filecolor=blue]{hyperref}
\usepackage{natbib}
\newcommand {\ctn}{\citet} 
\newcommand {\ctp}{\citep}

\newtheorem{definition}{Definition}[section]
\newtheorem{theorem}{Theorem}[section]
\newtheorem{lemma}[theorem]{Lemma}

\newtheorem{remark}[theorem]{Remark}

\newcommand{\abs}[1]{|#1|}
\newcommand{\norm}[1]{\Vert#1\Vert}
\newcommand{\R}{I\!\!R}

\newcommand{\bd}{\boldsymbol{d}}

\newcommand{\btheta}{\boldsymbol{\theta}}
\newcommand{\bLambda}{\boldsymbol{\Lambda}}

\newcommand{\bTheta}{\boldsymbol{\Theta}}

\newcommand{\bSigma}{\boldsymbol{\Sigma}}

\newcommand{\bmu}{\boldsymbol{\mu}}

\newcommand{\bD}{\boldsymbol{D}}

\newcommand{\bG}{\boldsymbol{G}}

\newcommand{\bk}{\boldsymbol{k}}

\newcommand{\bS}{\boldsymbol{S}}

\newcommand{\bR}{\boldsymbol{R}}

\newcommand{\bx}{\boldsymbol{x}}
\newcommand{\bX}{\boldsymbol{X}}

\newcommand{\bzero}{\boldsymbol{0}}
\newcommand{\bone}{\boldsymbol{1}}

\newcommand{\postp}{P_{\btheta|\bX_n}}
\newcommand{\pexp}{E_{\btheta|\bX_n}}

\newcommand{\dec}{\boldsymbol{d}}

\newcommand{\Rn}[1] {\R^{#1 }}
\DeclareMathOperator*{\argmin}{argmin}
\DeclareMathOperator*{\argmax}{argmax}
\DeclareMathOperator*{\ess}{ess}
\newcommand{\mfdr}{mFDR_{\bX_n}}
\newcommand{\fdrx}{FDR_{\bX_n}}
\newcommand{\mn}{\mathcal{MN}}
\begin{document}

\normalsize

\title{\vspace{-0.8in}
Non-marginal Decisions: A Novel Bayesian Multiple Testing Procedure}
\author{Noirrit K. Chandra and Sourabh Bhattacharya\thanks{
Noirrit K. Chandra is a PhD student and Sourabh Bhattacharya is an
Associate Professor in Interdisciplinary Statistical Research Unit, Indian Statistical Institute,
203, Barrackpore Trunk Road,
Kolkata - 700108, West Bengal, India. Corresponding e-mail: bhsourabh@gmail.com}}
\date{\vspace{-0.5in}}
\maketitle

\begin{abstract}
In this paper we consider the problem of multiple testing when the hypotheses are dependent. In most of the existing literature, either Bayesian or non-Bayesian, the decision rules mainly focus on the validity of the test procedure rather than actually utilizing the dependency to increase efficiency. Moreover, the decisions regarding different hypotheses are marginal in the sense that they do not depend upon each other directly.
However, in realistic situations, the hypotheses are usually dependent, and hence it is desirable that the 
decisions regarding the dependent hypotheses are taken jointly.

In this article we develop a novel Bayesian multiple testing procedure that coherently takes this requirement 
into consideration. Our method, which is based on new notions of error and non-error terms, substantially enhances efficiency 
by judicious exploitation of the dependence structure among the hypotheses. 
We prove that our method minimizes the posterior expected loss associated with a an additive ``0-1" loss function; we also prove 
theoretical results on the relevant error probabilities, establishing the coherence and usefulness of our method.
The optimal decision configuration is not available in closed form and we propose a novel and efficient 
simulated annealing algorithm for the purpose of optimization, which is also generically applicable to 
binary optimization problems. 

Numerical studies demonstrate that in dependent situations, our method performs significantly better than some 
existing popular conventional multiple testing methods, in terms of accuracy and power control. Moreover, application 
of our ideas to a real, spatial data set associated with 
radionuclide concentration in Rongelap islands yielded insightful results.
\\[2mm]
\end{abstract}

\section{Introduction}
\label{sec:intro}
In modern day practical statistical problems with many parameters we are seldom interested in testing only one hypothesis. 
Simultaneous inference on hundreds of parameters are often necessary, for instance, in spatial,  microarray datasets or in analysis of fMRI data. 
Thus, multiple testing has emerged as a very important problem in statistical inference. As in the case of single 
hypothesis testing with well-known notions of Type-I and Type-II errors, the multiple testing literature 
also consists of several measures of errors, for example, the family wise error rate ($FWER$), which is the probability 
of rejecting any null, the false discovery rate ($FDR$), which is the expected proportion of false discoveries, 
and false non-discovery rate ($FNR$), the expected proportion of false non-discoveries. \ctn{dudoit2003} discussed 
in details various issues related to controlling different types of errors. 

Several methods have been established to control different types of errors.
The $FWER$ controlling procedure uses the Bonferroni correction that rejects individual null hypotheses 
at $\alpha/m$ level of significance. This procedure is too conservative and results in low power for substantially 
large number of tests. \ctn{Benjamini95} proposed the powerful approach of controlling $FDR$. There have been much 
advancements both in the frequentist and Bayesian literatures for multiplicity correction later on. 
\ctn{berry99} have given a Bayesian perspective on multiple testing where the tests depend upon each other 
through a dependent prior. \ctn{scott06} discussed different aspects via a decision theoretic approach. 
Afterwards, \ctn{SanatGhosh08} introduced a general decision theoretic approach which controls the Bayes $FDR$ ($BFDR$) 
and Bayes $FNR$ ($BFNR$) criteria. In their paper, randomized decision rules have been introduced where the decisions 
of different hypotheses depend upon each other through a dependent structure. Dependence among test statistics 
naturally arises in many multiple testing scenarios. For example, in spatial data where the geographical 
locations are nearby, the test statistics for different hypotheses are quite likely to be influenced by each other. 
In microarray experiments, different genes may cluster into groups along biological pathways and exhibit high correlation. 
In public health studies, the observed data from different time periods and locations are often serially or 
spatially correlated. \ctn{by01} have shown that control over FDR 
is achieved for certain kinds of positive dependency among the tests. \ctn{finner2002, finner2007, efron2007} discussed the effect of 
dependence among test statistics, among others. \ctn{qiu2005} showed that dependence among test statistics 
significantly affects the power of many $FDR$ controlling procedures. \ctn{schwartzman2011} and \ctn{fan2012} 
discussed estimation of $FDR$ under correlation. 

However, in both classical and Bayesian literature, even in the dependent set-ups, 
most of the methods are concerned with marginal decision rules, 
in the sense that the decisions mainly depend upon the marginal distributions of the test statistics, 
marginal p-values or marginal posterior probabilities. In cases where we have additional information 
about dependency among tests, utilizing it will yield more reliable and closer-to-truth inference. 
Most of the methods focus on controlling the errors rather than actually utilizing the information 
supplied by the dependence structure. When two or more dependent hypotheses are being tested, 
decisions on different hypotheses are expected to influence each other. \ctn{sun2009} have discussed an  
approach for data arising from hidden Markov model, arguing that accounting for such dependence
increases efficiency. \ctn{caioptim11} have proposed 
an optimal decision rule for short range dependent data with dependent test statistics. However, both the works 
are about marginal decision rules, and their dependency is automatically accounted for in a Bayesian set-up. 
Indeed, in Bayesian multiple testing procedures, some implicit adjustment over multiplicity and dependence 
is naturally taken care of by considering dependent prior over the parameters, as the posterior distribution is 
influenced by the complete data. \ctn{scott10} discussed how empirical Bayes and fully Bayes methods adjust multiplicity.

When the decisions are not directly (deterministically) dependent, information provided by the 
joint structure {\it inherent in the hypotheses} are somewhat neglected by the marginal multiple testing approaches, even
though the data (and the prior in the Bayesian case) are dependently modelled. 
To illustrate, suppose that we want to test $H_{0i}:\theta_i\geq0$ vs. 
$H_{1i}:\theta_i<0$, $i=1,2$. Let $T_1$ and $T_2$ be the test statistics and suppose that they are highly positively 
correlated. Let us consider the decision rule that we reject $H_{0i}$ in favour of $H_{1i}$ if $T_i<c$, for some 
threshold $c$. Due to the high positive correlation between the test statistics, it is a natural guess that the 
tests should be accepted or rejected together. Suppose both the null hypotheses are true. However, for sampling perturbations, 
it is of course possible that $T_1<c$ but $T_2>c$, which would yield the counter-intuitive result that $H_{01}$ 
is rejected but $H_{02}$ is accepted. Using dependent decision rules should be helpful to rectify these kinds of errors 
if the information provided by the dependence is utilized judiciously. 
In this regard, in this paper we develop a novel multiple testing procedure that coherently takes the dependence
structure into consideration. 

Our procedure is based on new notions of error and non-error terms associated 
with breaking up the total number of hypotheses. We penalize the decision of each hypothesis by incorrect decisions regarding other dependent parameters. Thus we design a compound criterion where decisions regarding dependent parameters deterministically depend upon each other. We show that, by virtue of this dependent decision rule, our method asymptotically minimizes the Kullback-Leibler (KL) divergence from the true model. Also in extensive simulation studies with dependent data, we see that our method is quite advantageous in terms of the Type-II error.

We also propose a modified $FDR$ criterion where the dependence between parameters is incorporated in the error measure. We show that the modified version possess very desirable theoretical properties. Extensive simulation studies indicate that controlling the modified version provides extra safeguard by exploiting the dependence structure and results in lower Type-II error.

Moreover, we obtained insightful and interpretable results on application of our non-marginal method to a real, spatial example.

The rest of our paper is structured as follows. We introduce our non-marginal multiple testing procedure 
in Section \ref{sec:new_proposal} and develop new Bayesian error rates for our method in Section \ref{sec:mpBFDR}. In Section \ref{sec:optimality} we show optimality of 
the method with respect to appropriate loss functions based on the ``0-1" loss. In Section \ref{sec:KL_minimzation} we show that the non-marginal method minimizes the Kullback-Leibler divergence from the true model in an asymptotic sense. In Section \ref{sec:practical_issue} we discuss issues related to practical implementation of our multiple testing procedure. 
In this context, we propose and develop a novel simulated annealing algorithm 
for optimization of the criterion for our non-marginal method; this algorithm, however, is applicable to any
optimization problem consisting of binary variates.
We conduct simulation studies, demonstrating the superiority of our methods over some popular existing 
multiple testing methods in Section \ref{sec:simulation_study}, and in Section \ref{sec:realdata}, we
apply our ideas to a real spatial data set concerning radionuclide concentrations on Rongelap island.
Finally, we summarize our contributions and provide concluding remarks in Section \ref{sec:conclusion}.
The \textit{``S"} labelled equations and proofs of all our results are provided in the supplementary material.

\section{New proposal to obtain non-marginal decisions}
\label{sec:new_proposal}

\subsection{The basic multiple testing set-up }
\label{sec:basicsetup}
Let $\bX_n=(X_1, X_2,\ldots,X_n)$ be the observed data. Let the joint distribution of $\bX_n$ given $\btheta=(\theta_1,\theta_2,\ldots,\theta_m)$ be $P_{\bX_n|\btheta}(\cdot)$ where $\btheta$ are the parameters of interest and $\theta_i\in\varTheta_i$ for all $i=1,\ldots,m$. We put a prior $\Pi(\cdot)$ on the parameter space. Let $\postp(\cdot)$ and $\pexp(\cdot)$ be the posterior probability and posterior expectation of $\btheta$, respectively, given $\bX_n$. $P_{\bX_n}(\cdot)$ and $E_{\bX_n}(\cdot)$ 
represents the marginal distribution of $\bX_n$ and expectation with respect to this marginal distribution respectively.

Consider the following hypotheses:
$$ H_{0i}:\theta_i \in \varTheta_{0i}  \hbox{ vs. } H_{1i} : \theta_i \in \varTheta_{1i},  $$ 

where $\varTheta_{0i} \bigcap \varTheta_{1i}=\emptyset \mbox{ and } \varTheta_{0i} \bigcup \varTheta_{1i} 
= \varTheta_{i},\mbox{ for $i=1,\ldots,m$}.$

Here we discuss the multiple comparison problem in a Bayesian decision theoretic framework, given data $\bX_n$. 
For $i=1,\ldots,m$, let us first define the following quantities:
\begin{align*}
d_i=&\begin{cases}
1&\text{if the $i$-th hypothesis is rejected;}\\
0&\text{otherwise;}
\end{cases}\\
r_i=&\begin{cases}
1&\text{if $H_{1i}$ is true;}\\
0&\text{if $H_{0i}$ is true.} 
\end{cases}
\end{align*}

\ctn{muller04} considered the following additive loss function
\begin{equation}
L(\bd,\btheta)= c\sum_{i=1}^m d_i(1-r_i)+ \sum_{i=1}^m (1-d_i)r_i,
\label{eq:loss_mul}
\end{equation}
where $c$ is a positive constant.
The decision rule that minimizes the posterior risk of the above loss is given by:
\begin{equation}
d_i=I\left(v_i>\frac{c}{1+c}\right)~\text{for all }i=1,\cdots,m,
\label{eq:muldec}
\end{equation}
where $I(\cdot)$ is the indication function and $v_i=\postp(r_i)$.  
This loss function has been widely used in the Bayesian multiple testing literature and also in frequentist decision theoretic approaches. We consider these methods to be marginal because $d_i$ depends only on the marginal posterior probability of the $i^{th}$ hypothesis. 

In many real life situations auxiliary information regarding the dependence structure of the parameters are available. On the basis of such information suitable dependent prior distribution on 
the parameters is envisaged. For example in spatial statistics, Gaussian process prior is often considered. In fMRI data, Gaussian Markov random field prior is a common prior. In such cases, the additional information on the parameters are incorporated in the model through the prior distribution. Various applications in recent times in fields as diverse as spatio-temporal statistics, 
neurosciences, biological sciences, engineering, environmental and ecological sciences, astrostatistics, epidemiology, social sciences, psychometrics, demography, geostatistics, 
reliability engineering, statistical signal processing, statistical physics, finance, actuarial science, to name only a few, consider Bayesian analyses with dependent prior structures.
Our proposal is to incorporate such information, when available, in the testing procedure to obtain improved decision rule. This principle is in accordance with the traditonal Bayesian philosophy
that when prior information is available, inference can be enhanced. 
In this regard, we develop a multiple testing method where decisions regarding dependent hypotheses are not marginal as of (\ref{eq:muldec}). We elaborate our methodology in the next section.

\subsection{New error based criterion}
\label{sec:err}
Let $G_i$ be the set of hypotheses (including hypothesis $i$) where the parameters are 
dependent on $\theta_i$. Define the following quantity:
\begin{equation*}
z_i=\begin{cases}
1&\mbox{if $H_{d_j,j}$ is true for all $j\in G_i\setminus\{i\}$;}\\
0&\mbox{otherwise.}
\end{cases}
\end{equation*}
If $G_i$ is a singleton, then we set $z_i=1$.

Now consider the term 
\begin{equation}
TP=\sum_{i=1}^md_ir_iz_i. 
\label{eq:tp}
\end{equation}
This is the number of cases $i$ for which $d_i=1$, $r_i=1$ and $z_i=1$; in words, $TP$ is the number of cases for which the 
$i$-th decision correctly accepts $H_{1i}$, and all other decisions in $G_i$, which may accept either 
$H_{0j}$ or $H_{1j}$, for $j\neq i$, are correct. We refer to this quantity as the number of \textit{true positives}, 
and maximize its posterior expectation with respect to $\dec$. But there are also errors to be controlled, for example, 
\begin{align*}
E_1=&\sum_{i=1}^md_i(1-r_i)z_i;\\
E_2=&\sum_{i=1}^md_i(1-r_i)(1-z_i);\\
E_3=&\sum_{i=1}^md_ir_i(1-z_i).\\
\end{align*}
Here $E_1$ is the number of cases $i$ for which $d_i=1$, $r_i=0$ and $z_i=1$, that is, $E_1$ is the number of cases
for which $H_{1i}$ is wrongly accepted, but the remaining decisions in $G_i$ are correct;
$E_2$ is the number of cases for which $H_{1i}$ is wrongly accepted and at least one 
decision regarding the other hypotheses in $G_i$ is also wrong; $E_3$ is the number of cases for which
the $i$-th hypothesis is correctly rejected but at least one of the other decisions associated with $G_i$, is wrong.
The complete set of terms, corresponding to errors and correct decisions are provided in 
Section \ref{sec:appendix2}.
Adding up $E_1, E_2, E_3$ yields
\begin{equation}
E=\sum_{i=1}^md_i(1-r_iz_i),
\label{eq:err}
\end{equation}
which we will control, subject to maximizing $TP$. The following theorem, which we prove in Section \ref{subsec:proof_overpenalty} of the supplement, 
shows that controlling many error terms is not advisable.
Hence, we do not attempt to control the other error terms detailed in Section \ref{sec:appendix2}.
\begin{theorem}
	\label{th:overpenalty}
	Controlling many error terms while maximizing $TP$ in the non-marginal decision theoretic framework leads to over-penalization resulting in low power.
\end{theorem}

Note that $E$ is the total number of cases $i$ for which $d_i=1$, $r_iz_i=0$, that is, 
either the $i$-th hypothesis is wrongly rejected or some other decision(s) in $G_i$ is wrong, or both. 
This is regarded as the number of \textit{false positives} in our notion. Note that in the definitions of both $TP$ and $E$, $d_i$ is penalized by incorrect decisions in the same group. This forces the decisions to be jointly taken adjudging other dependent parameters. Taking decisions jointly have particular advantages over marginal decision rules. In Section \ref{sec:KL_minimzation}, we show that by virtue of the joint decision rule, the non-marginal procedure minimizes the KL-divergence from the true data-generating process. 

We will minimize the posterior expectation of $-TP$ given by \eqref{eq:tp} subject to controlling the 
posterior expectation of $E$. 
%
Hence, with $E$ to be controlled, the function to be minimized is given by
\begin{align}
g_\lambda(\dec) =& -\sum_{i=1}^{m}d_i \pexp[r_iz_i] + \lambda \sum_{i=1}^{m}d_i\pexp[1-r_iz_i]\nonumber\\
=& -\sum_{i=1}^{m}d_i w_i(\dec) + \lambda \sum_{i=1}^{m}d_i(1 - w_i(\dec) )\nonumber\\
=& -(1+\lambda)\sum_{i=1}^{m}d_i\left(w_i(\dec) - \frac{\lambda}{1+\lambda} \right)\label{},
\label{eq:fpre}
\end{align}
where
\begin{align}
w_i(\dec) = \pexp[r_iz_i]= \postp\left(H_{1i}\cap \{\underset{j\neq i, j\in G_i}\cap H_{d_j,j}\}\right).
\label{eq:w}
\end{align}
If $G_i$ is a singleton, then since $z_i=1$, we replace $w_i(\dec)$ with the marginal posterior 
probability $\postp\left(H_{1i}\right)$.

We will minimize $g_\lambda(\dec)$ with respect to $\dec$, or equivalently, we can maximize
\begin{equation}
\sum_{i=1}^{m}d_i\left(w_i(\dec) - \frac{\lambda}{1+\lambda} \right)=\sum_{i=1}^{m}d_i\left(w_i(\dec) - \beta \right)
=f_\beta(\dec),\mbox{ where }\beta=\frac{\lambda}{1+\lambda}.\label{eq:beta}
\end{equation}
\begin{definition}
	Let $\mathbb D$ be the set of all $m$-dimensional binary vectors denoting all possible decision configurations. Define $$\widehat{\bd}=\argmax_{\bd\in\mathbb{D}} f_\beta(\bd)$$ where $0<\beta<1$. Then $\widehat{\bd}$ is the \textit{optimal decision configuration} obtained as the solution of the non-marginal multiple testing method.
	\label{def:nmd}
\end{definition}
This $\beta$ is the penalizing constant balancing between $\pexp(TP)$ and $\pexp(E)$, and indeed plays the crucial role of balancing between Type-I and Type-II errors. This is formalized in 
Theorems \ref{theorem:theorem1} and \ref{theorem:theorem2}. 

There are several cluster-based approaches in the multiple testing literature. \ctn{heller06,benjamini2007} discussed a cluster-based analysis of fMRI data in the context of multiple testing; they  
formed clusters on the basis of correlations between different voxels. \label{pg}
Similarly in analysing spatial signals, \ctn{sun15} formed clusters consisting of spatial locations, 
and considered a single decision for each cluster. In these works, a whole cluster is regarded as a signal, that is, all the decisions regarding the parameters in a particular cluster are same.

On the other hand, in our methodology, the idea behind group formation is completely different from the idea of clustering in the aforementioned works. In our case, all the decisions within 
a group may not be same.
Decisions regarding hypotheses in a group highly influence each other through the $z_i$ term that we have introduced 
in Section \ref{sec:err}. Moreover, our groups are overlapping
in general, because of inter-dependence among hypotheses in different groups, and thanks to this, 
decisions in two different groups are also dependent. In all the aforementioned cluster based methods, decisions 
regarding different clusters are marginal. 
Such cluster based approaches are important in situations where signals appear in clusters. Our procedure is also applicable in such situations by forming groups of dependent clusters. In Section \ref{subsec:groups} we discuss how to form the groups in different contexts where the hypotheses are particularly dependent.

In Section \ref{sec:optimality} we show that  our proposed method is the optimal solution minimizing an additive ``0-1" loss function. When proper dependence 
structure between the hypotheses is present, the ``0-1" loss function is advocated by \ctn{Abram06}.
We further show that for sufficiently large sample size $n$, and under reasonable assumptions, the non-marginal method minimizes the Kullback-Leibler (KL) divergence from the true decisions; see Section \ref{sec:KL_minimzation}. In both the aforementioned contexts, joint decision making plays a crucial role and particularly enables minimization of KL divergence. 

It is important to observe that any multiple testing method concerning the loss function in (\ref{eq:loss_mul}) , for instance, can be viewed as a special case of our method where, 
for $i=1,\ldots,m$, $G_i=\{i\}$, that is, when we have no information about any dependence between the hypotheses. In situations, where no information regarding the dependence structure between parameters is available, our method boils down to the additive loss function based method.


In practical situations, where this method would be implemented to actually get the decision configuration, one needs to maximize  $f_\beta(\bd)$ with respect to $\bd$. A  simulated annealing algorithm is proposed in Algorithm \ref{algo:simulated_annealing} of Section \ref{sec:simulated_annealing} to carry out the maximization problem in practice. 
\section{New Bayesian error rates for our non-marginal procedure}
\label{sec:mpBFDR}
Before we introduce our notion of Bayesian error rates, we first provide a brief account of some 
existing classical and Bayesian error rates.

\subsection{A brief overview of error rates in multiple testing}
\label{subsec:error_rates}
\ctn{storey03} advocated the \textit{positive FDR} as a measure of Type-I error in multiple testing literature. The measure is defined as:
\begin{equation}
pFDR=E_{\bX_n} \left[ \sum_{\dec\in\mathbb{D}}  
\frac{\sum_{i=1}^{m}d_i(1-r_i)}{\sum_{i=1}^{m}d_i}\delta(\dec|\bX_n)\bigg{|}\delta(\bzero|\bX_n)=0 \right],
\label{eq:pfdr}
\end{equation}
where $\delta(\dec|\bX_n)$ is the probability of choosing the decision configuration $\dec$ 
according to the associated multiple testing procedure and $\bzero$ is the decision configuration that no null hypothesis is rejected. In case of non-randomized decision rules, 
$\delta(\dec|\bX_n)=1$ for the decision configuration which is chosen to be the final decision rule.

Under the prior distribution of $\btheta$, \ctn{SanatGhosh08} defined the posterior $FDR$ as
\begin{align}
posterior~FDR
=&\pexp \left[ \sum_{\dec\in\mathbb{D}}  
\frac{\sum_{i=1}^{m}d_i(1-r_i)}{\sum_{i=1}^{m}d_i\vee1}\delta(\dec|\bX_n) \right]\nonumber\\
=&\sum_{\dec\in\mathbb{D}}  \frac{\sum_{i=1}^{m}d_i(1-v_i)}{\sum_{i=1}^{m}d_i\vee1}\delta(\dec|\bX_n).
\label{eq:Psfdr}
\end{align}
Given data $\bX_n$, we denote the posterior $FDR$ by $\fdrx$. Now, the \textit{positive Bayesian FDR (pBFDR)} is the expectation of (\ref{eq:pfdr}) with respect to 
$\btheta$ or expectation of (\ref{eq:Psfdr}) with respect to the conditional distribution 
$[\bX_n|\delta(\dec=\bzero|\bX_n)=0]$, and is given by:
\begin{align*}
pBFDR &= E_{\bX_n} \left[ \sum_{\dec\in\mathbb{D}} 
\frac{\sum_{i=1}^{m}d_i(1-v_i)}{\sum_{i=1}^{m}d_i}\delta(\dec|\bX_n)\bigg{|}\delta(\bzero|\bX_n)=0 \right].
\end{align*}
The numerator term in $pFDR$ or $pBFDR$ is the number of false positives. $FDR$ is the expected proportion of 
false positives among all discoveries.

\subsection{A new Bayesian false discovery rate and its properties}
\label{subsec:mpBFDR}
In accordance with our new notion of false positives we modify the false discovery rate criteria. The posterior modified $FDR$ is defined by
\begin{align}
posterior~modified~FDR&=\pexp \left[ \sum_{\bd\in\mathbb{D}}  
\frac{\sum_{i=1}^{m}d_i(1-r_iz_i)} {\sum_{i=1}^{m}d_i\vee1} \delta(\bd|\bX_n) \right]\nonumber\\
&= \sum_{\bd\in\mathbb{D}}  \frac{\sum_{i=1}^{m}d_i(1-w_i(\bd) )} {\sum_{i=1}^{m}d_i\vee1} \delta(\bd|\bX_n).
\label{eq:mpfdr}
\end{align}

We call it as \textit{posterior} $mFDR$, in short $\mfdr$. Notably in this error rate, there are extra penalizations for incorrect decisions regarding dependent parameters in group. Though, this makes the error rate more conservative than the $FDR$, but it gives extra safeguard against Type-II error. \ctn{Chandra16} explicitly showed that the $\mfdr$ is directly associated to the deviation from the true distribution through its convergence rate. Taking expectation with respect to the marginal distribution of the data, we get the \textit{modified positive Bayesian} $FDR~(mpBFDR)$.
\begin{align}
mpBFDR 
&=E \left[ \sum_{\bd\in\mathbb{D}}  
\frac{\sum_{i=1}^{m}d_i(1-r_iz_i)}{\sum_{i=1}^{m}d_i}\delta(\bd|\bX_n)\bigg{|}\delta(\bzero|\bX_n)=0 \right] \nonumber\\
&= E_{\bX_n} \left[ \sum_{\dec\in\mathbb{D}}  
\frac{\sum_{i=1}^{m}d_i(1-w_i (\bd))}{\sum_{i=1}^{m}d_i}\delta(\bd|\bX_n)\bigg{|}\delta(\bzero|\bX_n)=0 \right] 
\label{eq:mbfdr}
\end{align}

One may speculate that the modified error criterion, as well as the decisions of the non-marginal procedure may change with different choices of groups. 
In this regard, we argue that group formation should be based on domain knowledge regarding the association between parameters. Notably, based on domain knowledge prior correlation structures are generally imposed on the parameters, so that given the prior, the groups remain fixed. 
In Section \ref{subsec:groups} we discuss several schemes of forming groups in different contexts, based on the prior knowledge. For different priors, the group structures
will of course be different, but then in that case the existing posterior $FDR$ or $pBFDR$ will be different as well. 
In fact the essence of Bayesian analysis lies in judicious 
choice of the prior. In other words, our method of selecting the groups as well as the Bayesian versions of $FDR$, which are all based on the prior structure, are coherent.

In Section \ref{subsec:validation}, extensive simulation studies show that by controlling the modified $FDR$ some existing popular multiple testing methods gain accuracy. 
This is not unexpected, given that the modified version is associated with a stricter penalty for incorrect decisions.

We have discussed in Section \ref{sec:basicsetup}, the additive loss function based marginal methods becomes a special case of our non-marginal procedure when $G_i=\{i\}$ for all $i=1,\cdots,m$. In that case both the modified versions boil down to their existing counterparts.
\subsection{Controlling $FDR$}
\label{subsubsec:mpBFDR control}
For our method we can control $mpBFDR$ exactly at any pre-specified level by properly choosing $\beta$. 
Under very minor assumptions we rigorously prove this in the following theorem.
\begin{theorem}
	\label{theorem:theorem1}
	Assume that for all $\beta\in (0,1)$, the events 
	$\left\{\bX_n:\sum_{i=1}^m d^\ast_iw_i(\bd^\ast)
	=\beta\sum_{i=1}^md^*_i\right\}$ 
	and 
	\\
	$\left\{\bX_n:\sum_{i=1}^m d_iw_i(\bd)-\sum_{i=1}^m d^\ast_iw_i(\bd^\ast)
	=\beta\left(\sum_{i=1}^md_i-\sum_{i=1}^md^*_i\right)\right\}$ 
	for two different decision configurations $\bd$ and $\bd^*$ in $\mathbb D$, have zero probabilities under $\bX_n$. Then
	$mpBFDR$ for the non-marginal procedure is continuous in $\beta$.
\end{theorem}
\begin{remark}
	\label{remark:remark1}
	By the above theorem, continuity of $mpBFDR$ with respect to $\beta$ clearly holds when $w_i(\dec)$ have continuous distributions
	(existence of density not necessary). Even for discrete distributions assigning zero probabilities
	to the sets 
	$$\left\{\bX_n:\sum_{i=1}^m d^\ast_iw_i(\dec^\ast)
	=\beta\sum_{i=1}^md^*_i\right\}$$ 
	and 
	$$\left\{\bX_n:\sum_{i=1}^m d_iw_i(\dec)-\sum_{i=1}^m d^\ast_iw_i(\dec^\ast)
	=\beta\left(\sum_{i=1}^md_i-\sum_{i=1}^md^*_i\right)\right\}$$ 
	for $\bd\in\mathbb D$,
	such continuity holds.  
\end{remark}

The importance of Theorem \ref{theorem:theorem1} is that it shows we can set the error measure exactly at any desired level through adjusting $\beta$; this would yield greater power than controlling the error with an upper bound. Observe that $\beta$ is the weight of the error defined in (\ref{eq:err}). This interpretation of $\beta$ as the penalizing factor between error and $TP$ becomes rigorous from the following lemma and theorem.

\begin{lemma}
	Let $\hat{\bd}=\argmax_{\bd\in\mathbb D} f_\beta(\bd)$. Then $\sum_{i=1}^m \hat d_i$ is decreasing in $\beta$.
	\label{lemma:discovery}
\end{lemma}
This lemma shows that $\beta$ penalizes the number of rejections, that is, with increasing $\beta$, the number of rejections decrease.
\begin{theorem}
	\label{theorem:theorem2}
	mpBFDR for the non-marginal procedure is non-increasing in $\beta$.
\end{theorem}

The continuity and non-increasing properties of $mpBFDR$ asserted by 
Theorems \ref{theorem:theorem1} and \ref{theorem:theorem2} together help us easily set the Type-I error at any
desired level. 

\begin{remark}
	Note that Theorem \ref{theorem:theorem1}, Lemma \ref{lemma:discovery} 
	and Theorem \ref{theorem:theorem2} hold without any restriction on the group structure. Since, as already discussed, the additive loss function based methods are special cases of our non-marginal
	procedure when dependence between the hypotheses is ignored, the above results are applicable to such marginal methods as well. In this regard, note that the constant $c$ in the additive 
	loss function acts as the penalizing constant between Type-I and Type-II errors in the marginal methods.
	\label{remark:marginal}
\end{remark}

\subsection{Type-II Errors in Multiple Testing}
\label{subsec:BFNR}

The \textit{positive False Non-Discovery Rate} is defined as
\begin{equation*}
pFNR=E_{\bX_n} \left[ \sum_{\dec\in\mathbb{D}}  
\frac{\sum_{i=1}^{m}(1-d_i)r_i}{\sum_{i=1}^{m}(1-d_i)}\delta(\dec|\bX_n)\bigg{|}\delta(\bone|\bX_n)=0 \right],
\end{equation*}
and the \textit{positive Bayesian False Non-Discovery Rate} is given by
\begin{align*}
pBFNR &= E \left[ \sum_{\dec\in\mathbb{D}} \frac{\sum_{i=1}^{m}(1-d_i)r_i}{\sum_{i=1}^{m}(1-d_i)}
\delta(\dec|\bX_n)\bigg{|}\delta(\bone|\bX_n)=0 \right]\\
&=E_{\bX_n} \left[\sum_{\dec\neq\bone}   
\frac{\sum_{i=1}^{m}(1-d_i)v_i}{\sum_{i=1}^{m}(1-d_i)}\delta(\dec|\bX_n)\bigg{|}\delta(\bone|\bX_n)=0 \right].
\end{align*}
where $\bone$ is the decision configuration that all the null hypotheses are rejected. Note that $pFNR$ and $pBFNR$ are the expected proportions of false non-discoveries among all non-discoveries. Therefore, these are regarded as measures of Type-II errors in the context of multiple testing. The simulation studies in Section \ref{sec:simulation_study} show that that the non-marginal method is quite advantageous in terms of incurring lower $pBFNR$ when compared to some existing methods.

\section{Optimality of the non-marginal method with respect to the ``0-1" loss function}
\label{sec:optimality}

The ``0-1" loss function in the multiple testing context is given by (see \ctn{Abram06}, for example): 
\begin{equation}
L\left(\bd^t,\bd\right)=\begin{cases}
0\mbox{ if}~\bd=\bd^t,\\
1 \mbox{ otherwise},
\end{cases}
\label{eq:zero_one_loss}
\end{equation}
where $\bd^t$ is the true decision configuration. Note that minimization of the posterior expected loss with respect to the above loss function is the same
as minimization of the posterior $w(\bd)=\postp\left(\cap_{i=1}^mH_{d_{i},i}\right)$
with respect to all possible decision configurations $\bd$.

In the next sections, we prove optimality of the non-marginal procedure keeping the number of discoveries fixed at some $k$. This $k$ can be looked upon as a parameter of the loss function in \eqref{eq:zero_one_loss}. Recall that the additive loss function defined in \eqref{eq:loss_mul} also has the parameter $c$. \ctn{Guindani09} showed that for the decision rule in \eqref{eq:muldec}, $pBFDR<1/(1+c)$. It is a general practice to choose $c$ such that the Type-I/Type-II error is controlled at some desired level. From Remark \ref{remark:marginal} we see that the number of discoveries and $pBFDR$ both decrease with increase in $c$. Hence, for a particular value of $c$, the number of discoveries also gets fixed, and choosing an appropriate $c$ is equivalent to fixing the number of discoveries in the additive 
loss function based approaches. The ``0-1" loss puts equal weight on the number of discoveries. To overcome this, \ctn{Abram06} put a prior on the number of discoveries and directly minimized the posterior risk of the ``0-1" loss function. We do not invoke this extra prior structure in our method and choose our penalizing constant $\beta$ subject to controlling Type-I error at some desired level or equivalently the number of discoveries. 

\subsection{Optimality when all the parameters are dependent}
\label{subsec:single_group1}
Let $\bG=\{G_1,\ldots,G_m\}$ denote any set of groups associated with the $m$ hypotheses. We first consider the case where all the parameters are dependent upon each other, that is, $G_i=\{1,\ldots,m\}$, for $i=1,\ldots,m$.
We show that for any arbitrary sample size, the non-marginal procedure is optimal with respect to the ``0-1" loss function. In other words, when $G_i=\{1,\ldots,m\}$, for $i=1,\ldots,m$, 
our non-marginal method is optimal among all multiple testing methods in the sense of minimizing 
the posterior risk of the ``0-1" loss subject to the same number of discoveries of the competing decision configurations.
We formalize this in the form of the following theorem.



\begin{theorem}
	\label{theorem:compare_methods}
	%
	Assume that for our non-marginal method, 
	$G_i=\{1,\ldots,m\}$, for $i=1,\ldots,m$. Then for any integer $k$ such that $0<k<m$,
	there exists
	$\hat\beta$ such that the corresponding decision output $\hat\bd$ 
	minimizes the posterior risk associated with the ``0-1" loss
	among all decisions $\bd^*$ satisfying $\sum_{i=1}^m d^*_{i}=k$. 
\end{theorem}

\subsection{Optimality in the case of block dependent parameters}
\label{subsec:multiple_groups}
In Section \ref{subsec:single_group1}, we have shown optimality of the non-marginal procedure where all the parameters are dependent. However, dependence among all the parameters may not be present always. In this section, we show that the non-marginal method is also optimal for block dependent parameters.

We assume that we have $s$ blocks, $r$-th block consisting of $o_r (=m_r-m_{r-1})$ dependent parameters, where $1\leq r\leq s$ and where $\sum_{r=1}^so_r=m$. We assume that the blocks do not possess any inter-dependence a priori. Therefore, for any parameter $\theta_j$ in $r$-th block, $G_j$ consists of all the parameters in that block. Clearly, 
there will be $s$ distinct groups which we denote by $\{G^*_1,\ldots,G^*_s\}$.

\begin{equation*}
\underbrace{1,\cdots,m_1}_{G_1^*}, \underbrace{m_1+1,\cdots,m_2}_{G_2^*},\cdots,\underbrace{  m_{s-1}+1, \cdots ,m_s}_{G_s^*}
\end{equation*}

Clearly, $G_i\neq\{1,\ldots,m\}$ for any $i=1,\ldots,m$ unlike the case in Section \ref{subsec:single_group1}. 
Now, for disjoint groups the ``0-1" loss function defined in (\ref{eq:zero_one_loss}) would be too restrictive. Therefore, for proper multiplicity control across all groups we define an additive ``0-1" loss function by levying appropriate weight over the blocks. For that purpose we first define the following quantities:
\begin{align}
&\bd_{G_r^*}=\left(d_{m_{r-1}+1},\ldots,d_{m_r} \right)^T,\notag\\
&k_{r}(\bd)=\sum_{i\in G^*_r}d_{i},\label{eq:defn}\\
&\tilde{\bk}(\bd)=(k_1(\bd),\ldots,k_s(\bd))^T,\notag\\
&\bS(k)=\left\{\tilde \bk=(\tilde k_{1},\ldots,\tilde k_{s})^T:\sum_{r=1}^s\tilde k_{r}=k\right\}.\notag
\end{align}
In the definition of $\bS(k)$, $\tilde k_{1},\ldots,\tilde k_{s}$  and $k$ are non-negative integers. 
Now, for any decision configuration $\bd$ such that $\sum d_i=k$ we have $\tilde{\bk}(\bd)\in\bS(k)$. Therefore, any $\tilde{\bk}\in\bS(k)$ corresponds to some decision configuration $\bd$ where $\tilde k_r$ is the number of discoveries in $G_r^*$. Now we define the partial loss function for $G_r^*$:
\begin{equation*}
L_r\left(\bd^t,\bd\right)=\begin{cases}
0\mbox{ if }\bd_{G_r^*}=\bd^t_{G_r^*},\\
1 \mbox{ otherwise},
\end{cases}
\end{equation*}
and hence the following additive loss function subject to the restriction that $\bd\in \bS(k) $:
\begin{equation}
L\left(\bd^t,\bd\right|k)=\min_{\tilde{\bk}\in\bS(k)} \sum_{r=1}^s\tilde{k}_r L_r\left(\bd^t,\bd\right).
\label{eq:additive01}
\end{equation}
In the above additive loss function, each partial loss-function $L_r$ is weighted proportional to the number of discoveries in $G_r^*$ and then adjusting the weights such that the total loss is minimum. Optimality of the non-marginal based method is formalized in the following theorem:
\begin{theorem}
	\label{theorem:compare_methods2}
	%
	Assume that the parameters are block dependent a priori. 
	Then 
	the decision output $\hat\bd$ of the non-marginal based method
	minimizes the posterior risk associated with $L(\bd^t,\bd|k)$ subject to $\sum_{i=1}^m\hat d_i=k$, for any integer k where $1\leq k\leq m$. 
\end{theorem}

\subsection{Interpretation of posterior $mFDR$ as appropriate probabilities}
\label{subsubsec:mpBFDR_probability}
Note that given $\bX_n$, and assuming that $\bG=\{G_1,\ldots,G_m\}$ with $G_i=\{1,\ldots,m\}$ for $i=1,\ldots,m$, 
the $\mfdr$ 
boils down to $1-w(\hat\bd)$ by virtue of
Lemma \ref{lemma:equal_posteriors}, where
$\hat\bd$ is the decision configuration output of the non-marginal method. In this case $\mfdr$ is the posterior probability of the joint decision being wrong.


Now, consider the set-up in Section \ref{subsec:multiple_groups}. In this case, 
$$\mfdr=\sum_{\bd\in\mathbb D}\frac{\sum_{i=1}^md_{in}(1-w_i(\bd))}{\sum_{i=1}^md_{i}}\delta(\dec|\bX_n)
=\sum_{r=1}^s\frac{k_r(\hat\bd)}{\sum_{i=1}^m\hat d_{i}}w_{r}(\hat\bd^c_{G^*_r}).$$ 

Note that $w_{r}(\hat\bd^c_{G^*_r})=1-w_{r}(\bd_{G^*_r})$ is the probability that at least one decision in $r$-th block is incorrect. Recall that in the additive ``0-1" loss function defined in (\ref{eq:additive01}), $k_r(\bd)$ is the weight of the partial loss incurred in the $r$-th block, that is in $G_r^*$, for all $r=1,\ldots,s$. Similarly in $\mfdr$, the ratio $k_{r}(\hat\bd)/\sum_{i=1}^m\hat d_{i}$ can be interpreted as weight 
of the error incurred in group $G^*_r$. \ctn{genovese06} discussed weighted false discovery control and also proposed a way to estimate the weights corresponding to each hypothesis. In their method, the estimated weights were also proportional to how strong the signal was for each hypothesis. Similar ideology works behind interpreting $k_{r}(\hat\bd)/\sum_{i=1}^m\hat d_{i}$ as the weight 
for $G_r^*$. This probability, being proportional to the number of discoveries associated with group $G^*_r$, can be interpreted as the strength of group $G^*_r$ with respect to the number of discoveries associated with it.

Moreover, the weights add up to 1 and therefore, it is natural to think of them as probabilities. We  interpret the weight as the probability of occurring error in the corresponding block given the data. Thus, we see that
\begin{align*}
&\frac{\sum_{r=1}^sk_{r}(\hat\bd)(1-w_{r}(\bd_{G^*_r}))}{\sum_{i=1}^m\hat d_{i}}=\sum_{r=1}^sP(G^*_r|\bX_n)w_{r}(\hat\bd^c_{G^*_r})\\
&=\sum_{r=1}^sP(G^*_r|\bX_n)P(\hat\bd^c_{G^*_r}|\bX_n,G^*_r)
=\sum_{r=1}^sP(\hat\bd^c_{G^*_r},G^*_r|\bX_n)=P\left(\cup_{r=1}^s\left\{\hat\bd^c_{G^*_r}\cap G^*_r\right\}|\bX_n\right),
\end{align*}
which is the posterior probability that at least one decision in one of the blocks is incorrect.
Hence, in this case also $\mfdr$ can be interpreted as an appropriate probability.


We remark that for Bayesian multiple testing methods, in keeping with the Bayesian philosophy, it makes sense to define the error measures conditional on the data, avoiding expectation
with respect to the (marginal) distribution of the data. Not only does this support the Bayesian philosophy, it also drastically simplifies the computation of such error
measures in complex practical problems. 
Moreover, it can be easily verified that all the desirable properties of $mpBFDR$ remain intact even without the expectation with respect to the marginal distribution of data. 
It follows that the {\it bona fide} Bayesian version of $mpBFDR$ admits the interpretation as a valid posterior probability with all desirable properties under suitable assumptions.


\section{Minimization of the Kullback-Leibler divergence of the non-marginal multiple testing procedure}
\label{sec:KL_minimzation}
In this section, we show that the non-marginal method minimizes the KL-divergence from the true model. \ctn{Shalizi09} provided sufficient conditions for posterior convergence under general dependence set-up and showed. We briefly state the relevant results in the following section. 
\subsection{Preliminaries for ensuring posterior convergence under general set-up}
\label{sec:shalizi}
We consider a probability space $(\Omega,\mathcal F, P)$, 
and a sequence of random variables $X_1,X_2,\ldots$,   
taking values in some measurable space $(\Xi,\mathcal X)$, whose
infinite-dimensional distribution is $P$. 
We denote the distributions of the class of proposed models 
by $P_{\bX_n|\btheta}$, where $\btheta$ is associated with a measurable
space $(\bTheta,\mathcal T)$.
For the sake of convenience, we assume, as in Shalizi, that $P$
and all the $P_{\bX_n|\btheta}$ are dominated by a common reference measure, with respective
densities $p$ and $f_{\btheta}$. The usual assumptions that $P\in\bTheta$ or even $P$ lies in the support 
of the prior on $\bTheta$, are not required for Shalizi's result, rendering it very general indeed. We levy the prior distribution $\pi(\cdot)$ on the parameter space $\bTheta$. 
Consider the following likelihood ratio:
\begin{align*}
R_n(\btheta)=\frac{f_{\btheta}(\bX_n)}{p(\bX_n)}.
\end{align*}
For every $\btheta\in\Theta$, the KL-divergence rate $h(\btheta)$ is defined as 
\begin{equation*}
h(\btheta)=\underset{n\rightarrow\infty}{\lim}~\frac{1}{n}E\left(\log\frac{p(\bX_n)}{f_{\btheta}(\bX_n)}\right),
\end{equation*}
given that the above limit exists. For $A\subseteq\bTheta$, let
\begin{align}
h\left(A\right)=\underset{\btheta\in A}{\mbox{ess~inf}}~h(\btheta);~ 
J(\btheta)=h(\btheta)-h(\Theta);~
J(A)=\underset{\btheta\in A}{\mbox{ess~inf}}~J(\btheta).\label{eq:J2}
\end{align}

We state assumptions \ref{shalizi1}--\ref{shalizi7} considered by Shalizi in Section \ref{subsec:assumptions_shalizi} of the Appendix. 
Under those assumptions the following theorem can be seen to hold:
\begin{theorem}[\ctp{Shalizi09}]
	\label{th:shalizi}
	Consider assumptions \ref{shalizi1}--\ref{shalizi7} and any set $A\in\mathcal T$ with $\pi(A)>0$. If $\varsigma>2h(A)$, where
	$\varsigma$ is given in (\ref{eq:S5_1}) under assumption \ref{s5}, then
	\begin{equation*}
	\underset{n\rightarrow\infty}{\lim}~\frac{1}{n}\log\postp(A|\bX_n)=-J(A).
	\end{equation*}
\end{theorem}

\subsection{KL-divergence when all the parameters are dependent}
Let $\bG=\{G_1,\ldots,G_m\}$ denote any set of groups associated with the $m$ hypotheses. We consider the case where all the parameters are dependent upon each other as in Section \ref{subsec:single_group1}, that is, $G_i=\{1,\ldots,m\}$, for $i=1,\ldots,m$. Note that, the possible decision configurations corresponding to the $m$ hypotheses partitions the parameter space $\bTheta$ into $2^m$ partitions. Let $\bTheta_{\bd}=\{\theta_1\in\Theta_{d_1},\ldots,\theta_m\in\Theta_{d_m}  \}$. From \eqref{eq:J2}, we see that $J(\bTheta_{\bd})$ is the $\ess\inf$ KL-divergence rate from the true model. Now, we state the following theorem:
\begin{theorem}
	\label{th:single_group}
	Assume that for our non-marginal method, 
	$G_i=\{1,\ldots,m\}$, for $i=1,\ldots,m$. Then for any integer $k$ such that $0<k<m$,
	there exists
	$\hat\beta$ such that the corresponding decision output $\hat\bd$ 
	asymptotically minimizes the KL divergence rate $J(\bTheta_{\bd})$ among all decision configurations $\bd$ satisfying $\sum_{i=1}^m d_{i}=k$. 
\end{theorem}
\subsection{KL-divergence minimization in case of block-dependent parameters}
\label{subsec:multiple_group}
Similar to Section \ref{subsec:multiple_groups}, we now consider the case where the parameters are block dependent.
$\bG=\{G_1,\ldots,G_m\}$ be the set of all groups and 
$\{G^*_1,\ldots,G^*_s\}$ denote the set of distinct and disjoint groups, where $1<s\leq m$. 

As in Section \ref{subsubsec:mpBFDR_probability} we also assume that the hypotheses within the groups $G^*_i;$ $i=1,\ldots,s$, correspond to 
parameter sets $\Theta^*_r;$ $r=1,\ldots,s$, and that these parameter sets are associated with independent data sets. 
In other words, we assume that $\bX_n=\{\bX_{1n},\ldots,\bX_{sn}\}$ and the likelihood is of the form $\prod_{r=1}^s[\bX_{rn}|\Theta^*_r]$.


Let us now consider the problem of maximization of 
$$\sum_{i=1}^md_{i}\left(w_{in}(\bd)-\beta_n\right)
=\sum_{r=1}^s\left(\sum_{i\in G^*_r}d_{i}\right)\left(w_{rn}(\bd_{G^*_r})-\beta\right)
=\sum_{r=1}^sk_{r}(\bd)\left(w_{rn}(\bd_{G^*_r})-\beta\right),$$ 
subject to $\sum_{i=1}^md_{i}=k$, as in Section \ref{subsec:multiple_groups}.
With the way of maximization of the individual summands $k_{r}(\bd_{G^*_r})\left(w_{rn}(\bd_{G^*_r})-\beta\right)$ for fixed $k_{r}(\bd_{G^*_r})=\tilde k_{r}$ as detailed in 
Section \ref{subsec:multiple_groups}, it is clear that the maximization problem is equivalent to maximization of 
$\frac{1}{mn}\sum_{r=1}^sk_{r}(\bd_{G^*_r})\left(\log w_{rn}(\bd_{G^*_r})-\beta^*\right)$
for fixed $\tilde k_{r}$; $r=1,\ldots,s$, where we have replaced $w_{rn}(\bd_{G^*_r})$ by $\log w_{rn}(\bd_{G^*_r})$ and $\beta^*=\log\beta$.

Now let the groups $\left\{G^*_1,\ldots,G^*_s\right\}$ be homogeneous in the sense that 
\begin{equation}
\underset{m\rightarrow\infty}{\lim}~\frac{k_r(\bd^t)}{m}=p\in (0,s^{-1}), 
\label{eq:homo}
\end{equation}
where
$\bd^t$ is the true decision configuration. In words, for large number of hypotheses $m$, the proportion of true discoveries are approximately the same for all the groups $G^*_r$.
Let $\tilde k_r=k_r(\bd^t)$.
Then subject to 
\begin{equation}
\underset{m\rightarrow\infty}{\lim}~\frac{k_r(\bd)}{m}=\underset{m\rightarrow\infty}{\lim}~\frac{\tilde k_r}{m}=p;~r=1,\ldots,s, 
\label{eq:constraint1}
\end{equation}
let us consider maximization of
\begin{equation}
\underset{m\rightarrow\infty}{\lim}~\frac{1}{mn}\sum_{r=1}^sk_{r}(\bd)\left(\log w_{rn}(\bd_{G^*_r})-\beta^*\right)
\label{eq:nmd_infinite}
\end{equation}
with respect to $\bd$. 
Let $\bTheta_{\bd^{\infty}}$ be the parameter space associated with the infinite dimensional decision configuration $\bd^{\infty}$. Then the following result holds.
\begin{theorem}
	\label{th:multiple_group}
	Assume the above set-up of disjoint and distinct groups $\{G^*_1,\ldots,G^*_s\}$ satisfying the homogeneity condition (\ref{eq:homo}). Then 
	there exists $\hat\beta$ such that the corresponding non-marginal decision output $\hat\bd^{\infty}$ maximizing (\ref{eq:nmd_infinite})
	asymptotically minimizes the KL divergence rate $J(\bTheta_{\bd^{\infty}})$ among all decision configurations $\bd^{\infty}$ satisfying (\ref{eq:constraint1}). 
\end{theorem}

\section{Practical issues on implementation of the non-marginal procedure}
\label{sec:practical_issue}
Performance of the non-marginal procedure heavily depends upon the choice of groups as the decisions significantly depend upon each other through the group structure. 
Judicious choice of the groups is thus crucial for our methodology. Also, proper choice of the penalization constant $\beta$ plays a major role in the procedure as does the constant 
$c$ in the additive loss function defined in (\ref{eq:loss_mul}). And finally, the problem of obtaining the optimal decision configuration by maximizing $f_\beta(\bd)$ must receive
its due attention. Indeed, since the decisions deterministically depend upon each other, 
the decision rule is not available in closed form, and sophisticated numerical methods must be employed to optimize $f_\beta(\bd)$. In this section we discuss the solutions in details.

\subsection{Choice of $\left\{G_1,\ldots,G_m\right\}$}
\label{subsec:groups}
From the Bayesian perspective, we recommend the choice of $\left\{G_1,\ldots,G_m\right\}$ using the prior correlation structure between the parameters of interest. In cluster-based multiple testing approaches, \ctn{benjamini2007} prescribed formation of the clusters using information outside the data to be analysed. Therefore, from the Bayesian viewpoint, their recommendation seems to coincide with our idea of forming groups on the basis of prior correlation. 


Recall that $G_i$ is defined as the set of parameters with inherent dependence structure with $\theta_i$. However, in implementation of the method forming groups concerning all dependent parameters might be disadvantageous in high dimensional cases.


Firstly, keeping very weakly dependent parameters in $G_i$ will only increase the complexity of the method without rendering any extra information from the dependent structure. 
This can be explained heuristically as follows. 
Recall from Definition \ref{def:nmd} that the quantity $\sum_{i=1}^md_i(w_i(\bd)-\beta)$ is maximized with respect to $\bd$. Now, the joint posterior probability $w_i(\bd)$ will tend to be small (often, less than $\beta$) if $G_i$ is consisted of numerous parameters. Keeping very weakly dependent parameters in the group will incur over-penalization levying high posterior probability of $z_i=0$. In such cases, the decision configuration $\bd$, with $d_i=0$ for all $i$, will tend to be the solution of 
the maximization problem if even a single decision in the same group is incorrect. This might turn the method to be overly conservative.

A second disadvantage of large groups is related to the curse of dimensionality in computing the high dimensional joint posterior probabilities $w_i(\bd)$. For large $m$, the numerical values of this probability will often be quite small, again prompting unreasonable selection of many null hypotheses.

Both the problems are avoided if the group sizes chosen are not significantly large. We provide the following scheme of group formation on the basis of prior dependence structure. \ctn{Chandra16} showed that the non-marginal method is robust on group formation in the sense that it asymptotically converges to the true decisions.

Assume that the prior correlation structure between the $m$ parameters
of interest is given by $\bR^m$ with $(i,j)$-th element $\rho_{ij}$. We first consider the correlations between 
the $i$-th and $j$-th parameters, with $i<j$, 
and obtain the desired percentile (say, 95\%) $\rho$ of these quantities. 
Then, in $G_i$ we include only those indices $j~(\neq i)$ such that $\rho_{ij}\geq\rho$.
Thus, the $i$-th group contains indices of the parameters that are highly correlated with the $i$-th parameter. If there exists no index $j$ such that $\rho_{ij}\geq\rho$, then $G_i=\{i\}$. This scheme of group formation has yielded excellent results in the simulation studies in Section \ref{sec:simulation_study}.

In some special cases this group formation strategy can be further simplified and is often complimentary to the situation. While testing for spatial signals, groups can be formed with neighbouring locations in each group. This strategy is implemented in Section \ref{sec:realdata} and elaborately explained in Section \ref{subsubsec:groups_realdata}.

In functional Magnetic Resonance Imaging (fMRI) studies, multiple testing is commonly used to detect actual signals and separating out noise. \ctn{zhang2011} proposed a methodology of local aggregation of voxels, subsequently applying to a multiple testing method. Also in the Bayesian approach to fMRI studies, the Markov-random field (MRF) prior has been widely used in the literature. 
Since MRF considers dependence structure among the neighbouring voxels, our method of group formation using dependent neighbouring sites (voxels)
is in keeping with the local dependence structure induced by the MRF prior. Thus, the idea of forming groups in this manner is parallel to the general strategy of group formation on the basis of prior correlation that we have already discussed.

In microarray or microRNA datasets, multiple testing is widely used to detect differentially expressed genes. Information are available on positional and functional clustering of genes. 
Incorporating these information in the model as prior and subsequently forming groups might help account for the dependence between the genes and yield better results.

In situations, where no prior information on dependence structure is available, groups can be formed on the basis of the dependence structure showcased by the data. 
This is similar to the empirical Bayes procedures of prior selection. 


\subsection{Choice of the penalizing constant $\beta$}
\label{subsec:beta_choice}
In Section \ref{subsubsec:mpBFDR_probability}, we advocate the $\mfdr$ as a measure of Type-I error in multiple testing. Let $\hat{\bd}=\argmax_{\bd\in\mathbb{D} } f_\beta(\bd)$. 
We define $\gamma(\beta)=\frac{\sum_{i=1}^m\hat d_i(1-w_i(\hat\bd)) }{\sum_{i=1}^m\hat d_i}$. Clearly, $\gamma(\beta)$ is the $\mfdr$ incurred given the data. Then
\[
\gamma(\beta)<1-\beta.
\]

\begin{figure}[H]
	\centering
	\includegraphics[trim={0 .25in .4in .8in },clip, totalheight=0.2\textheight]{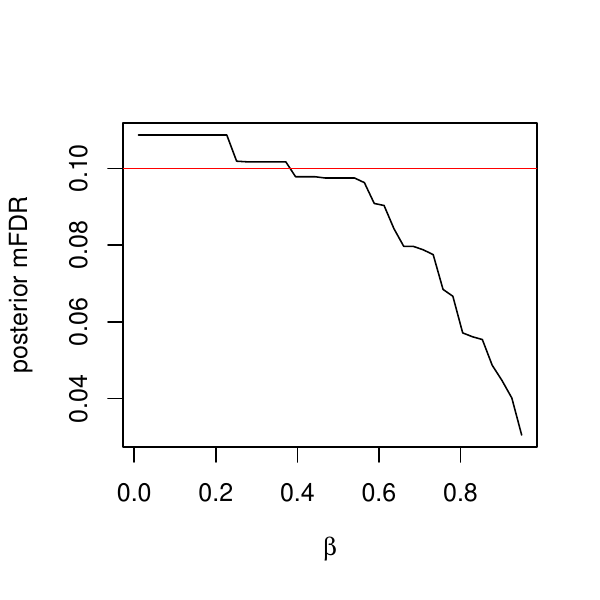}
	\caption{}
	\label{fig:choose_beta}
\end{figure}

However, considering this property only might lead to very conservative control. To illustrate, suppose that one is interested in controlling the error at level 0.1. 
Figure \ref{fig:choose_beta} shows that considering $\beta=0.9$ would wield a conservative decision configuration where the actual error is much lower than 0.1. 

From Theorem \ref{theorem:theorem2} we see that the error is non-increasing in $\beta$. In light of this theorem, we propose the following algorithm to choose $\beta$, assuming
that the interest lies in controlling $\gamma(\beta)$ at level $\alpha$. 
\begin{algorithm}
	\caption{Algorithm to choose appropriate $\beta$}
	\label{algo:choose_beta}
	\algdef{SE}[DOWHILE]{Do}{doWhile}{\algorithmicdo}[1]{\algorithmicwhile\ #1}%
	
	\begin{algorithmic}[1]
		\State Start with $\beta=1-\alpha$, compute $\mathbb{E}=\gamma(\beta)$ and take a small $\epsilon>0$.
		\While {$\mathbb{E}\leq\alpha$}\\
		Set $\tilde{\beta}=\beta-\epsilon$ and compute
		$\mathbb{E}=\gamma(\tilde\beta)$.\\
		Set $\beta=\tilde{\beta}$. 
		\EndWhile	
		\State \textbf{end while}.
		\State $\hat{\beta}=\beta+\epsilon$ is the appropriate value of $\beta$.
		
	\end{algorithmic}	
\end{algorithm}


\subsection{A novel simulated annealing methodology for optimization with binary variables and application to our decision problem}
\label{sec:simulated_annealing}

In this section, we propose a novel and efficient \textit{simulated annealing} methodology 
to solve the penalized optimization problem of maximizing $f_\beta(\bd)$. Importantly, this method is applicable to all optimization problems involving any number of binary variables. 


Simulated annealing is an MCMC based stochastic optimizing algorithm that is capable of escaping the attraction
of the local modes. 
Because of the ease of implementation and particularly thanks to the ability to escape local modes, 
this algorithm is particularly very useful for optimizing 
arbitrary complicated functions with many local modes. 
The fundamental idea, discussed in \ctn{Robert13}, is a change of scale, 
called \textit{temperature}, allowing faster moves on the surface of the function to be maximized. 
This rescaling partially avoids the possibility to get stuck in a local maximum. Given a temperature parameter $T_i>0$, 
sample is generated from 
\begin{equation*}
\pi_i(\dec)\propto \exp\{ f_\beta(\dec) \times T_i\}.
\end{equation*}
As $T_i$ increases towards infinity, the values simulated from this distribution become concentrated 
in narrower and narrower neighbourhoods around the global maxima of $f_\beta$.

Samples are generated from $\pi_i(\cdot)$ by the Metropolis-Hastings (MH) strategy. 
At each step, the simulation method perturbs the values of the variables by a small amount, while $T_i$ is slowly increased simultaneously. 
The resultant configuration will be accepted if it procures a higher value of $f_\beta(\cdot)$. If not, then also the new configuration
can be accepted with a positive acceptance probability. This enables the system 
to hill-climb from a locally optimal state. 
Simulated annealing is the repeated application of the above basic step until no more increment 
of the desired function is virtually possible.
This method always assigns positive probability to the event of escaping a local maximum. 

However, in multiple testing 
contexts, hundreds and thousands and sometimes even millions of hypotheses 
are tested simultaneously making $\bd$ quite high-dimensional. In the case using an ordinary MH-based algorithms to generate sample from $\pi_i(\cdot)$ often becomes inefficient affecting the acceptance rate and convergence of simulated annealing. \ctn{Somak14} devised a Transformation based MCMC (TMCMC) method and showed that even though a very large number of parameters are to be updated, these can be updated 
very efficiently by simple deterministic transformations of a single, one-dimensional random variable with high acceptance rate. 
Therefore to generate samples from $\pi_i(\cdot)$ in the simulated annealing algorithm, we implement the TMCMC strategy. 

Note that, in the problem of maximizing $f_\beta(\cdot)$, each component of $\dec$ is either 0 or 1, that is, the support of $f_\beta$ is a finite set with discrete binary vectors. 
In Algorithm \ref{algo:simulated_annealing}, we describe the TMCMC based simulated annealing algorithm for the optimization problem. We random update the component(s) of the vector with the univariate quantity $\xi$. This algorithm can be applied in complex discrete optimization problems.

\begin{algorithm}
	\caption{Maximization of $f_\beta$ by \textit{Simulated Annealing} using $TMCMC$}
	\label{algo:simulated_annealing}
	
	
	\begin{algorithmic}[1]
		\State Start with $\dec^{(0)}= (d_1^{(0)},\cdots,d_m^{(0)})$, and set $\xi=1$. Fix some probability
		$r\in (0,1)$.
		\For{$i=0\cdots N$} 
		\State $\dec^\ast=\dec^{(i)}$ and simulate $v\sim U(0,1)$. 
		\If{$v<r$} change exactly one of $d_j^\ast$'s, $j=1,\cdots,m$. Decide randomly which one to update. 
		If, say, $d_1^\ast$ is the selection, then set $d_1^\ast=(d_1^\ast+\xi)\mod 2$ and 
		$d_j^\ast$ is unchanged $\forall\:j=2,\cdots,m$.
		
		\Else \:update all $d_j^\ast$'s. Set $d_j^\ast=(d_j^{(i)}+\xi)\mod 2\: \forall\:j=1,\cdots,m$.
		\EndIf
		\State Simulate $u\sim U(0,1)$.
		
		\begin{enumerate}[label=\roman*]
			\If{$u<\alpha=\min\left\lbrace1,\frac{\pi_i(\dec^\ast)}{\pi_i(\dec)} \right\rbrace=\min\left\lbrace1, e^{(f_\beta(\dec^\ast)-f_\beta(\dec) )\times T_i} \right\rbrace  $} update $\dec^{(i+1)} =\dec^\ast$.
			\Else \:$\dec^{(i+1)}=\dec^{(i)}$.
			\EndIf 
		\end{enumerate} 
		\EndFor
		
		
	\end{algorithmic}
	
\end{algorithm}

The decision configuration $\dec$, obtained by running the above algorithm for a sufficiently large number
of iterations, 
is the optimal decision configuration maximizing $f_\beta(\dec)$. Note that, in steps 4 and 5 of the algorithm, a new proposal 
value is generated. Effectively by simulating only one sample from the uniform distribution on $(0,1)$, we are able 
to generate this proposal value in arbitrary dimensions. This saves huge computational cost when the 
dimension is reasonably high. 
Note that updating only one co-ordinate randomly with probability $r$ and updating all the co-ordinates with the remaining probability is required for irreducibility which we prove subsequently.


In our simulated annealing algorithm, the time inhomogeneous Markov chain that has been used, 
has the following transition kernel:
\begin{align}
P_i(\dec^\ast|\dec)&=\mathbb{P}(\text{at step $i+1$ the decision config is }\dec^\ast|
\text{at step $i$ the decision config was }\dec)\notag \\
&= \mathbb{P}(S^{(i+1)}=\dec^\ast|S^{(i)}=\dec).
\label{eq:transkern}
\end{align}

Simple verification of the detailed balance condition leads to the following lemma. 
\begin{lemma}	
	The Markov chain with transition kernel (\ref{eq:transkern}) has stationary distribution $\pi_i$ and it is also irreducible and aperiodic.
	\label{lemma:lemmastationary}
\end{lemma}


As is well-known, in practice, judicious choice of the temperature is crucial for satisfactory convergence
of simulated annealing algorithms. In our simulation studies and the real data example, the choice
$T_i=\log\log(i+2);~i\geq 1$, turned out to be very appropriate in this regard. In all our applications,
we considered $10^6$ iterations and chose that decision configuration for which the optimizing criterion is the
maximum among $10^6$ iterations. 

\section{Simulation study}
\label{sec:simulation_study}
In this section we compare the performance of the non-marginal procedure $(NMD)$ with the following widely used methods of \ctn{muller04} ($MPR$), \ctn{SanatGhosh08} ($SZG$), \ctn{Benjamini95} ($BH$) and \ctn{storey02} ($ST$) respectively. We elaborate the simulation design in the following section.


\subsection{The true data generating mechanism}
\label{subsec:true_model}
Let $$\bX_1,\bX_2,\ldots,\bX_n\overset{iid}{\sim}\mn(\bmu^m,\bSigma^m),$$ where $\mn(\bmu^m,\bSigma^m)$ stands for
multivariate normal with mean $\bmu^m$ and dispersion matrix $\bSigma^m$. In this simulation experiments $\bSigma^m$ is a non-diagonal positive-definite matrix constructed in the following manner. We set $\bSigma^m=\bD^m\bR^m{\bD^m}^T$; here $\bD^m$ is an $m$-dimensional diagonal matrix where the diagonal elements are distributed independently and identically as the square root of the chi-square distribution with five degrees of freedom. We assume that the $(i,j)$-the element of $\bR^m$ is of the form $\exp\left\{-(z_i-z_j)^2\right\}$, where $z_1,\ldots,z_m\stackrel{iid}{\sim} Beta\left(\frac{1}{2},\frac{1}{2}\right)$.
The hypotheses of our interest are
\begin{equation}
H_{0i}:\mu_i \geq 0  \hbox{ vs. } H_{1i} : \mu_i <0,~i=1,\ldots,m.\label{eq:hyp}
\end{equation}

Consider a scalar value $a\in [-1,1]$. Now for each replication of our simulation experiment, we first simulate $\bmu^m$ from $\mn(a\bone^m,\bSigma^m)$ 
and treat these simulated values of $\bmu^m$ as true values. 
Then we draw the sample $\bX_1,\bX_2,\ldots,\bX_n$ from $\mn(\bmu^m,\bSigma^m)$, for the particular value of $a$.
Once the data is thus generated with a dependence structure, we compare the performance of the non-marginal method with the competing methods across the replications.

We perform the experiments for 21 equispaced values of $a$ in $[-1,1]$. This $a$ can be regarded as a shift parameter from true null to true alternative. Note that when $a$ is close to $1$, most of the nulls will happen to be true. Similarly most of the nulls will be false when $a$ is close to $-1$. 
Though the latter case is not very practical in real life situations, for the sake of completeness we nevertheless perform the simulation studies for comparing the performance of different methods.

We have done 1500 replications of the simulation experiment for all our subsequent studies. Once generated $\bSigma_m$ is kept fixed throughout all our replications. 

\subsection{The postulated Bayesian model and $p$-value computation}
\label{subsec:model}
Note that our proposed $NMD$ method, and the competing $MPR$ and $SZG$ methods are Bayesian methods of multiple testing. All these methods require the posterior distribution of the parameters to carry out the hypothesis testing problem in (\ref{eq:hyp}). We state the likelihood and prior distribution considered for all these methods as following. We assume
\begin{equation}
\bX_1,\bX_2,\ldots,\bX_n\overset{iid}{\sim}\mn(\bmu^m,\bLambda^m).\label{eq:data}
\end{equation}

Notably, $(\bmu^m,\bLambda^m)$ are unknown parameters. We assume \textit{Normal-Inverse Wishart } $(NIW)$ prior on these parameters.
\begin{equation}
(\bmu^m,\bLambda^m)\sim NIW(a\bone^m,\lambda, \bSigma^m,\nu),\label{eq:niw}
\end{equation}
where $a$ is the shift parameter mentioned in Section \ref{subsec:true_model}, $\lambda=1$ and $\nu=m$. Then the posterior distribution of $(\bmu^m,\bLambda^m)$ given data $(\bX_1,\bX_2,\ldots,\bX_n)$ is $$\left[\bmu^m,\bLambda^m|\bX_1,\bX_2,\ldots,\bX_n\right]\sim NIW (\tilde{\bmu},\tilde\lambda,\tilde{ \bLambda},\tilde\nu),$$ 
where
\[
\tilde\nu=\nu+n,\tilde\lambda=n+\lambda, \tilde{\bmu}=\frac{n\bar{\bX}+\lambda a\bone}{n+\lambda},\tilde{\bLambda}
=n\bS+\bSigma^m+\frac{n\lambda}{n+\lambda}(\bar{\bX}-a\bone)(\bar{\bX}-a\bone)^T.\\
\]
In the above,
$\bar{\bX}$ and $\bS$ are the sample mean and dispersion matrix, respectively. In (\ref{eq:niw}), integrating out $\bLambda^m$ we get
\begin{equation}
\left[\bmu^m|\bX_1,\bX_2,\cdots,\bX_n\right]\sim t_{\tilde\nu-m+1} \left( \tilde{\bmu},\frac{1}{\tilde\lambda(\tilde\nu-m+1)}\tilde{\bLambda}\right),\label{eq:post_mu}
\end{equation}

where $t_\nu(\bmu,\bLambda)$ denotes multivariate central $t$-distribution with location vector $\bmu$, scale matrix $\bLambda$ and $\nu$ degrees of freedom. All the three Bayesian methods, namely $NMD$, $MPR$ and $GMZ$ are performed with respect to the posterior distribution of $\bmu^m$ in (\ref{eq:post_mu}). Also the $NMD$ method requires to define the groups. We have implemented the group formation strategy discussed in Section \ref{subsec:groups} in our simultation studies on the basis of the prior correlation matrix $\bR^m$.



As regards $BH$ and $ST$, these are frequentist methods. Due to normality of the data (see  (\ref{eq:data})), the hypothesis testing problem in (\ref{eq:hyp}) is equivalent to testing 
$H_{0i}:\mu_i =0$ vs. $H_{1i} : \mu_i <0$, by virtue of the monotone likelihood ratio property. The Student's $t$-test statistic which is also the \textit{most powerful} test statistic in this case, is given by $T_i=\frac{\sqrt{n}\bar{X}_i}{s_i}$ where $\bar{X}_i$ and $s_i$ are the sample mean and standard deviation respectively. Clearly under $H_{0i}$, $T_i$ follows a $t$-distribution with $n-1$ degrees of freedom for all $i$. The $p$-value corresponding to $i$-th test is given by $p_i=P(t_{n-1}<T_i)$. The $BH$ method is executed on the basis of these $p$-values subject to controlling the $FDR$ at the required level.

However, it is not straightforward to compare with the method $ST$. As in that method, $FDR$ is estimated 
for a fixed rejection region, whereas we set the rejection region subject to controlling Type-I error at a fixed level. In this method, $i$-th null hypothesis is rejected if $p_i<\Gamma$ and $FDR$ is computed for that particular $\Gamma$.
We circumvent this problem by setting $\Gamma$ such that the $FDR$ is controlled at the requisite level. 
\subsection{Comparison Scheme for Performance Comparison to Competing Methods}
To compare the performance of our NMD method with the competing Bayesian methods, we control versions of $FDR$ at the same level for all the methods and study the respective $pBFNR$ incurred.



However, for any frequentist method, the Bayesian error rates $mpBFDR$ or $pBFDR$ are undesirable
since these measures are prior-dependent although the methods are not. Therefore, we consider $mpFDR$ and $pFDR$ for our purpose which are Monte Carlo averages of the quantities
$\frac{\sum_{i=1}^md_i(1-r_iz_i)}{\sum_{i=1}^md_i}$ and $\frac{\sum_{i=1}^md_i(1-r_i)}{\sum_{i=1}^md_i}$
over the simulation replicates. Note that, in simulation studies, 
$r_i$; $i=1,\ldots,m$, are known, so that it is straightforward to compute the above quantities.

\subsection{Validation of $mpBFDR$}
\label{subsec:validation}

Since versions of $FDR$ play significant roles in multiplicity control, it is important
to select the appropriate version, particularly when comparing different multiple testing methods. 
Hence, before conducting the simulation study for such comparison, we first consider selection of suitable 
versions of false discovery rates.

We recommend to control the modified version of $FDR$ proposed by us. As already discussed this measure has extra penalization for incorrect decisions regarding other dependent parameters also. This provides extra safeguard against incurring Type-II error apart from controlling the Type-I error. In this section, we provide evidence towards our claim 
through simulation studies. We demonstrate that controlling the modified versions of $FDR$ leads to closer to truth inference for existing marginal multiple-testing methods also.

For each of the methods $MPR$ and $SZG$, we compute the proportion of making correct decisions regarding all the hypotheses. We name it \textit{proportion of true decision (PTD)}. For the aforementioned Bayesian methods, we compute $PTD$ against controlling $mpBFDR$ and $pBFDR$ separately. The $pBFDR$ is controlled at level $0.05$ and $mpBFDR$ is controlled at the minimum level achieved by the respective methods.

However, for large number of hypotheses, it is practically impossible to obtain correct decisions for every hypothesis. 
So, in a pathological example with 3 hypotheses we demonstrate that controlling $mpBFDR$ yields larger
$PTD$. As such, we conduct a simulation study with the true data generated from the mechanism described in
Section \ref{subsec:true_model} and the model proposed in Section \ref{subsec:model}, with $a=0$, $m=3$ and $n=10$. 
Here we consider 
$G_i=\{1,2,3\},i=1,2,3$, that is, we consider the complete dependent structure of all the parameters. The simulation results are summarized in Table \ref{table:tab1}. These results indicate that even for the marginal Bayesian methods $MPR$ and $SZG$, controlling $mpBFDR$ is advantageous and yields better inference.

\begin{table}[h]
	\caption{Versions of $FDR$ control in Bayesian methods}
	\label{table:tab1}
	\centering
	\begin{tabular}{ll}
		\begin{tabular}{|c|c|}
			\multicolumn{2}{c}{$MPR$}\\\cline{1-2}
			Type-I error rate	&  $PTD$\\ \cline{1-2}
			$mpBFDR=0.2675$	&  0.7653\\ 
			$pBFDR=0.05$	& 0.7220\\\cline{1-2}
		\end{tabular} 
		&
		\begin{tabular}{|c|c|}
			\multicolumn{2}{c}{$SZG$}\\\cline{1-2}
			Type-I error rate	&  $PTD$\\ \cline{1-2}
			$mpBFDR=0.2558$	&  0.6793\\ 
			$pBFDR=0.05$	& 0.6673\\\cline{1-2}
		\end{tabular}
	\end{tabular}
\end{table}




As already discussed, $mpBFDR$ or $pBFDR$ are prior based error measures, which are not appropriate
for frequentist methods. Hence, in this case we consider $mpFDR$ and $pFDR$. Here also we compute $PTD$ while controlling $pFDR$ and $mpFDR$ separately. $pFDR$ is controlled at level 0.05 and $mpFDR$ is controlled at the minimum level achieved by the methods. The results are summarized in Table \ref{table:tab2}. the results re-iterate that even for the frequentist marginal multiple testing methods, controlling the modified versions of $FDR$ is advantageous. 

\begin{table}
	\caption{Versions of $FDR$ control in frequentist methods}
	\label{table:tab2}
	\centering
	\begin{tabular}{ll}
		\begin{tabular}{|c|c|}
			\multicolumn{2}{c}{$BH$}\\\cline{1-2}
			Type-I error rate	&  $PTD$\\ \cline{1-2}
			$mpFDR=0.2151$ & 0.7747\\ 
			$pBFDR=0.05$	& 0.7420\\\cline{1-2}
		\end{tabular} 
		&
		\begin{tabular}{|c|c|}
			\multicolumn{2}{c}{$ST$}\\\cline{1-2}
			Type-I error rate	&  $PTD$\\ \cline{1-2}
			$mpFDR=0.2625$ & 0.7673 \\ 
			$pBFDR=0.05$	& 0.7393\\\cline{1-2}
		\end{tabular}
	\end{tabular}
\end{table}

Thus from Tables \ref{table:tab1}-\ref{table:tab2} we see that, incorporating the information regarding dependence structure in the error measure is important even for marginal methods. Even for very small sample size, controlling the modified $FDR$ leads to more accurate inference. 
In the following section, we compare the performance of our $NMD$ method with the competing methods 
with large number of hypotheses.

\subsection{Comparison of performances of multiple testing methods in terms of $pBFNR$ and $pFNR$}
\label{subsec:performance_comparison}
In Section \ref{subsec:validation}, we see that by controlling the modified version of the $FDR$, more accurate results are obtained from the existing marginal methods. In this section we study the Type-II error incurred by the $NMD$ method compared to the competing methods while controlling the modified $FDR$ at the same level. The set-up is as described in Section \ref{subsec:true_model} and \ref{subsec:model}. For the $NMD$ method, the groups are formed following the strategy discussed in Section \ref{subsec:groups}, with
the covariance structure given by $\bSigma^m$. We have taken $m=160$ and $n=20$ in this simulation study.

\begin{figure}[h]
	\centering
	\subfloat[][] {{\includegraphics[trim={0 .25in .4in .8in },clip, totalheight=0.275\textheight]{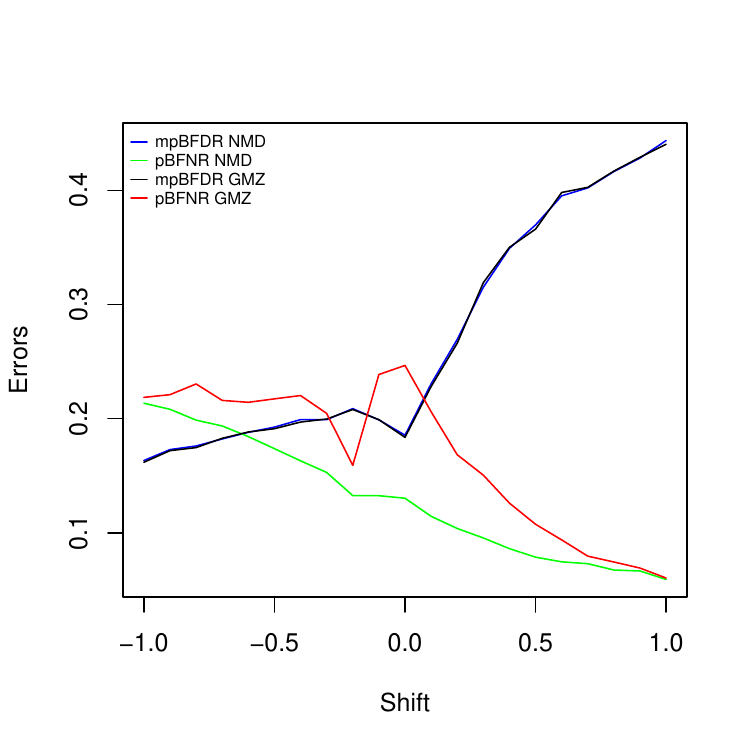}}\label{MPR}}	
	\subfloat[][]{{\includegraphics[trim={0 .25in .4in .8in },clip, totalheight=0.275\textheight]{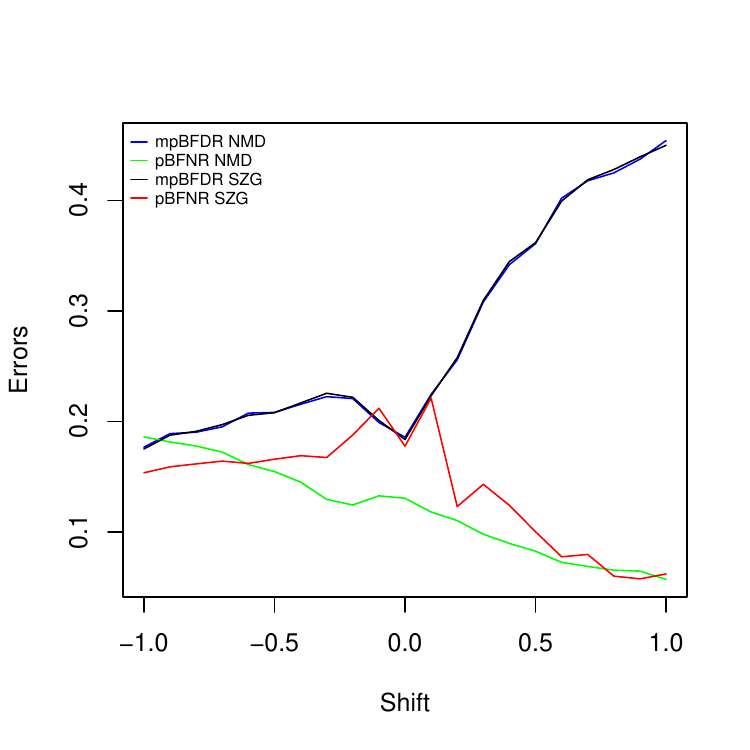}}\label{SZG}}
	\caption{$FNR$ comparison with Bayesian methods: \protect\subref{MPR} $MPR$	
		\protect\subref{SZG} $SZG$ }
	\label{fig:compare_Bayesian_methods}
\end{figure}

For the Bayesian methods, we compare $pBFNR$ while setting $mpBFDR$ at a fixed level. 
For each of the two competing Bayesian methods, we control $mpBFDR$ approximately at the minimum level they could achieve and then compare the $pBFNR$ incurred with that of the $NMD$ method for different values of the \textit{shift} parameter. 
The results depicted in 
Figure \ref{fig:compare_Bayesian_methods} show that compared to both the competing Bayesian methods, our $NMD$ 
method has incurred significantly and almost uniformly lesser Type-II error.

\begin{figure}
	\subfloat[][] { { \includegraphics[trim={0 .25in .4in .8in },clip, totalheight=0.275\textheight]{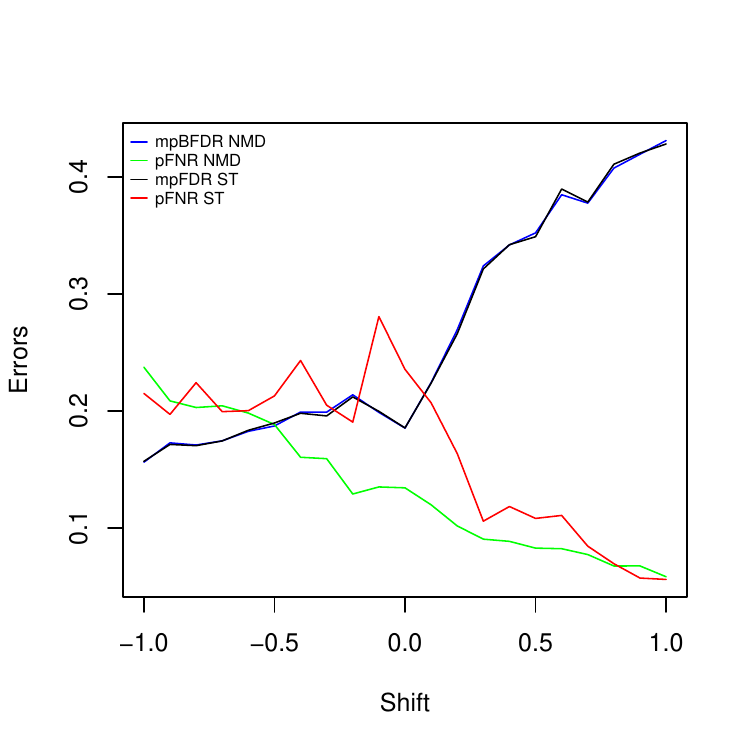}}\label{st}}	
	\subfloat[][]{ { \includegraphics[trim={0 .25in .4in .8in },clip, totalheight=0.275\textheight]{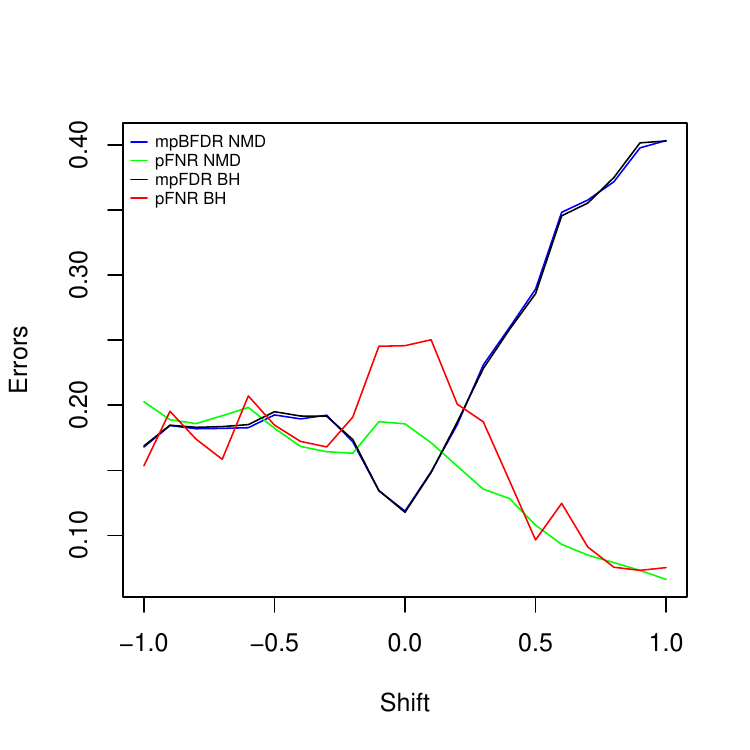}}\label{bh}}
	\caption{$FNR$ comparison with frequentist methods: \protect\subref{st} $ST$	
		\protect\subref{bh} $BH$. 
	}
	\label{fig:compare_classical_methods}
	
\end{figure}

For comparison of our method with the frequentist multiple testing methods $BH$ and $ST$, we control
$mpBFDR$ for our method and $mpFDR$ for the competing methods, setting the two error rates to be approximately equal, and compare the respective $pFNR$ incurred by the methods. Figure \ref{fig:compare_classical_methods} 
shows that the $NMD$ method incurred lesser $FNR$ compared to the competing frequentist methods also.


\begin{figure}[h]
	\centering
	\subfloat[][] { { \includegraphics[trim={0 .25in .4in .8in },clip, totalheight=0.275\textheight]{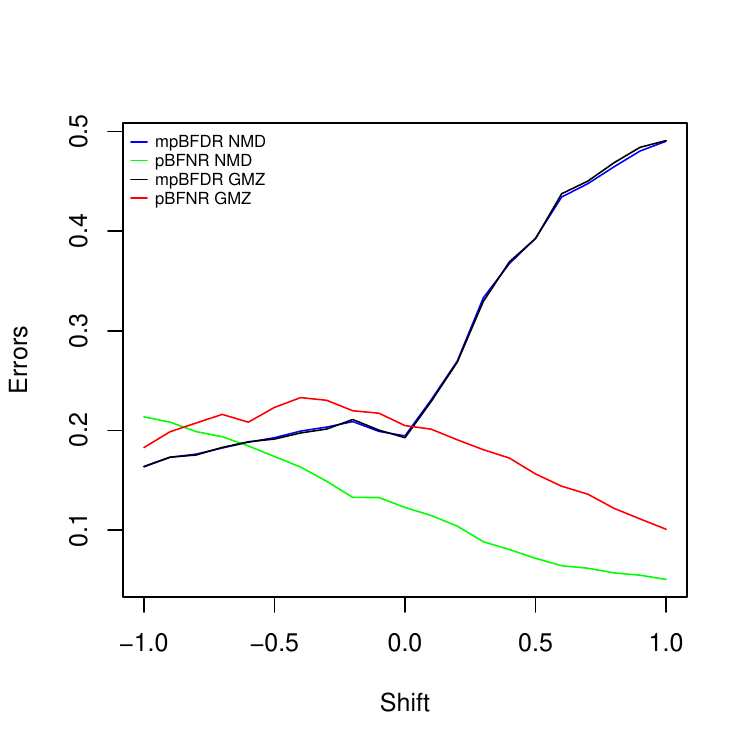}}\label{MPR5}}	
	\subfloat[][]{ { \includegraphics[trim={0 .25in .4in .8in },clip, totalheight=0.275\textheight]{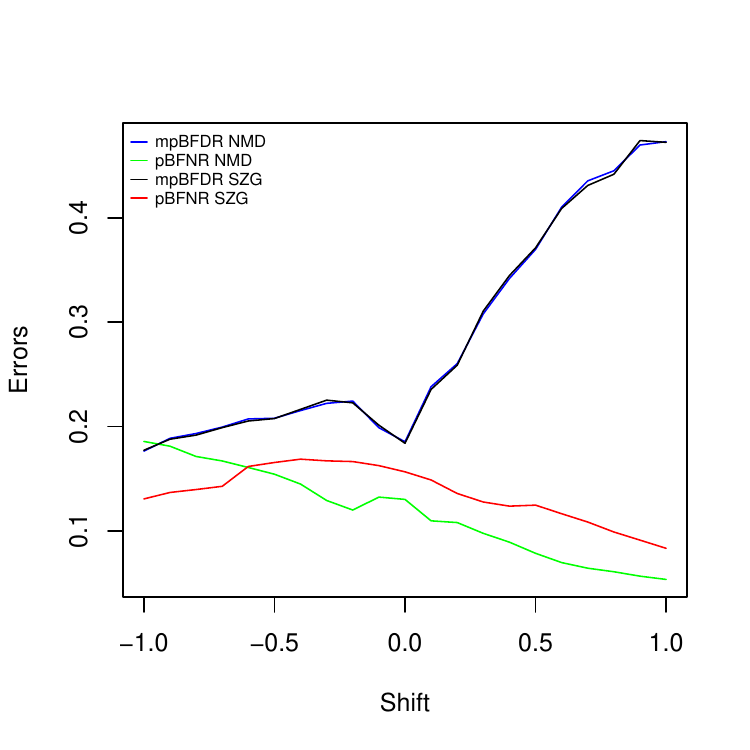}}\label{SZG5}}
	\caption{$FNR$ comparison with Bayesian methods: \protect\subref{MPR5} $MPR$	
		\protect\subref{SZG5} $SZG$ }
	\label{fig:compare_Bayesian_methods5}
\end{figure}

\begin{figure}	
	\subfloat[][] { { \includegraphics[trim={0 .25in .4in .8in },clip, totalheight=0.275\textheight]{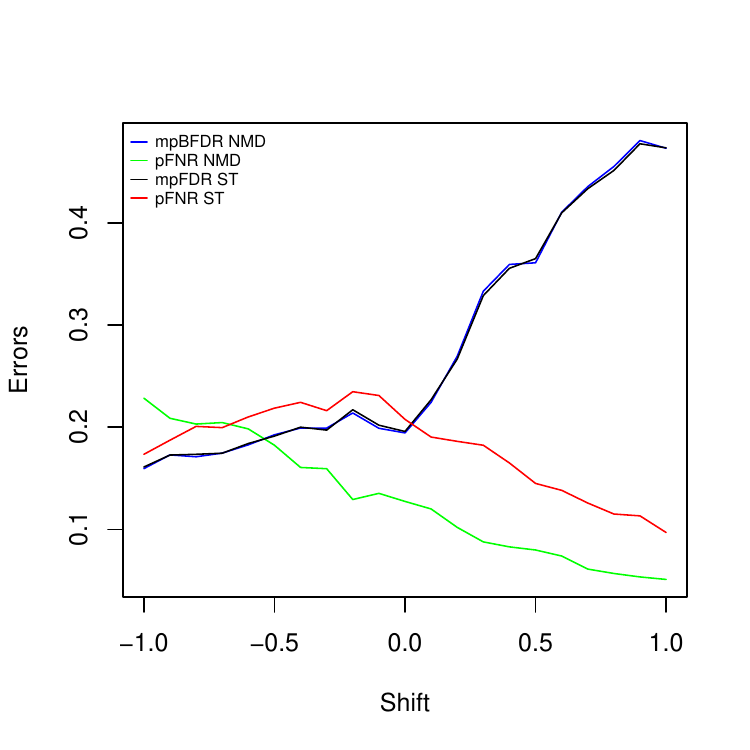}}\label{st5}}	
	\subfloat[][]{ { \includegraphics[trim={0 .25in .4in .8in },clip, totalheight=0.275\textheight]{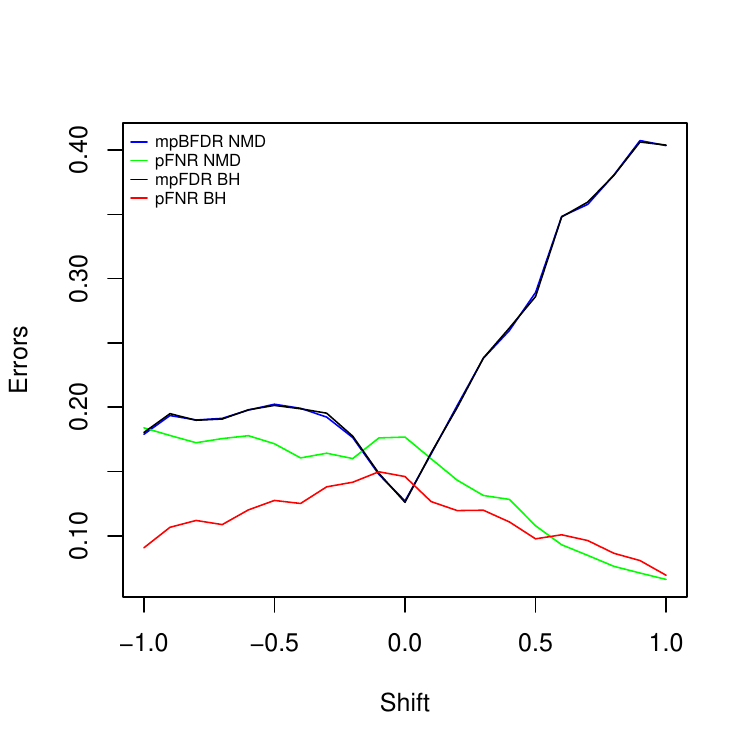}}\label{bh5}}
	\caption{$FNR$ comparison with frequentist methods: \protect\subref{st5} $ST$	
		\protect\subref{bh5} $BH$ }
	\label{fig:compare_classical_methods5}
\end{figure}

However, the competing methods are not designed to control the modified versions of $pBFDR$ or $pFDR$. Therefore, we have conducted another experiment controlling $pBFDR$ and $pFDR$ at level 0.05 
for the Bayesian and frequentist methods, respectively. For the Bayesian $MPR$ and $SZG$ methods, the incurred $mpBFDR$ is estimated when $pBFDR$ controlled at 0.05. Then the $NMD$ method is performed and compared to the two methods controlling the $mpBFDR$ at the respective estimated levels. For the frequentist methods, similar comparisons are done by controlling the $mpFDR$.
The comparisons are shown in Figures \ref{fig:compare_Bayesian_methods5} and \ref{fig:compare_classical_methods5}.

In Figures \ref{fig:compare_Bayesian_methods}-\ref{fig:compare_classical_methods5}, 
it is interesting to note that for values of the shift $a$ close to $-1$, 
$FNR$ incurred by the $NMD$ method is slightly higher in comparison. This admits the following explanation. Recall the data generating scheme in Section \ref{subsec:true_model} and observe that for values of $a$ close to $-1$, most of
the null hypotheses are false. In Section \ref{subsec:groups} we pointed out that in order to avoid over-penalization, the group sizes in the $NMD$ method should be chosen to be moderate.
Even though restricting the group sizes significantly mitigates
the problem of too much bias towards accepting the null hypotheses, the problem is not entirely eliminated, and plays some role when most of the nulls are false. In this case, the advantage of borrowing strength from dependence among the hypotheses is overridden by the extra penalization. 
As such, when the shift is close to $-1$, that is, when most of the null hypotheses are false,
the $NMD$ method is expected to have slightly lesser power. This is also reflected in the figures. However, in practice, based on expertise and domain knowledge null hypotheses are generally chosen such that most of them are expected to be true. Therefore, the situation where most nulls are false is practically unrealistic. Nevertheless, we conduct simulation experiments with $a$ close to $-1$ to compare the performance of the $NMD$ method with the others and see that the performances are quite comparable.
On the other hand, for larger values of $a$, which is the case in most practical applications, the dependence among
the hypotheses is adequately exploited by the non-marginal method to obtain much better performance. 
Among the competing methods, the BH procedure
deserves special mention. Indeed, Figure \ref{bh5} shows that the BH method performs better than the non-marginal procedure for the values of $a$ less than or equal
to about $0.6$ when $pFDR$ is controlled at level $0.05$; the non-marginal method begins to gain superiority only for $a$ larger than $0.6$, when most of the nulls begin to be true, 
which is a somewhat favourable situation for the non-marginal procedure. Figure \ref{bh} shows that the BH method
is not very easily outperformed by the non-marginal method even when $mpFDR$ is controlled. Since the actual data are positively correlated and since the BH procedure works well
under positive dependence \ctp{by01}, the above observations may possibly admit some explanation in this light.

As an aside, observe that in Figures \ref{fig:compare_Bayesian_methods}, \ref{fig:compare_classical_methods}, \ref{fig:compare_Bayesian_methods5} and \ref{fig:compare_classical_methods5},  
$mpBFDR$ and $mFDR$ are increasing with
$a$. We explain this phenomenon as follows. Note that as $a$ approaches $+1$, the proportion of true nulls also increases, 
giving room for falsely rejecting more true null hypotheses. 
Since no version of $FDR$ takes this information into account, 
the available versions of Type-I error in multiple testing also increases with $a$. 

\section{Real data analysis: radionuclide concentrations at Rongelap Atoll}
\label{sec:realdata}
Rongelap Atoll is a coral atoll of 61 islands in the Pacific Ocean, and forms a legislative district of the 
Ralik Chain of the Marshall Islands. On March 1, 1954, the United States conducted a nuclear test on Bikini Atoll 
in the northern Marshall Islands code named Bravo that led to widespread fallout contamination over inhabited islands 
of Rongelap, Ailinginae, and Utrok Atolls. Prior to Bravo, little consideration was given to the potential health 
and ecological impacts of fallout contamination beyond the immediate vicinity of the test sites. People living on Rongelap Atoll received significant exposure to ``fresh" radioactive fallout and had to be evacuated 
to Kwajalein Atoll for medical treatment. The Rongelap community spent the next 3 years living on Ejit Island 
(Majuro Atoll) before returning home to Rongelap in June 1957. However, growing concerns about possible long-term 
health effects associated with exposure to residual fallout contamination on the island prompted residents to 
relocate again to a new temporary home on Mejatto Island on Kwajalein Atoll in 1985.

\begin{figure}[h]
	\begin{center}
		\includegraphics[trim={.775in .85in .57in .95in },clip, totalheight=0.25\textheight]{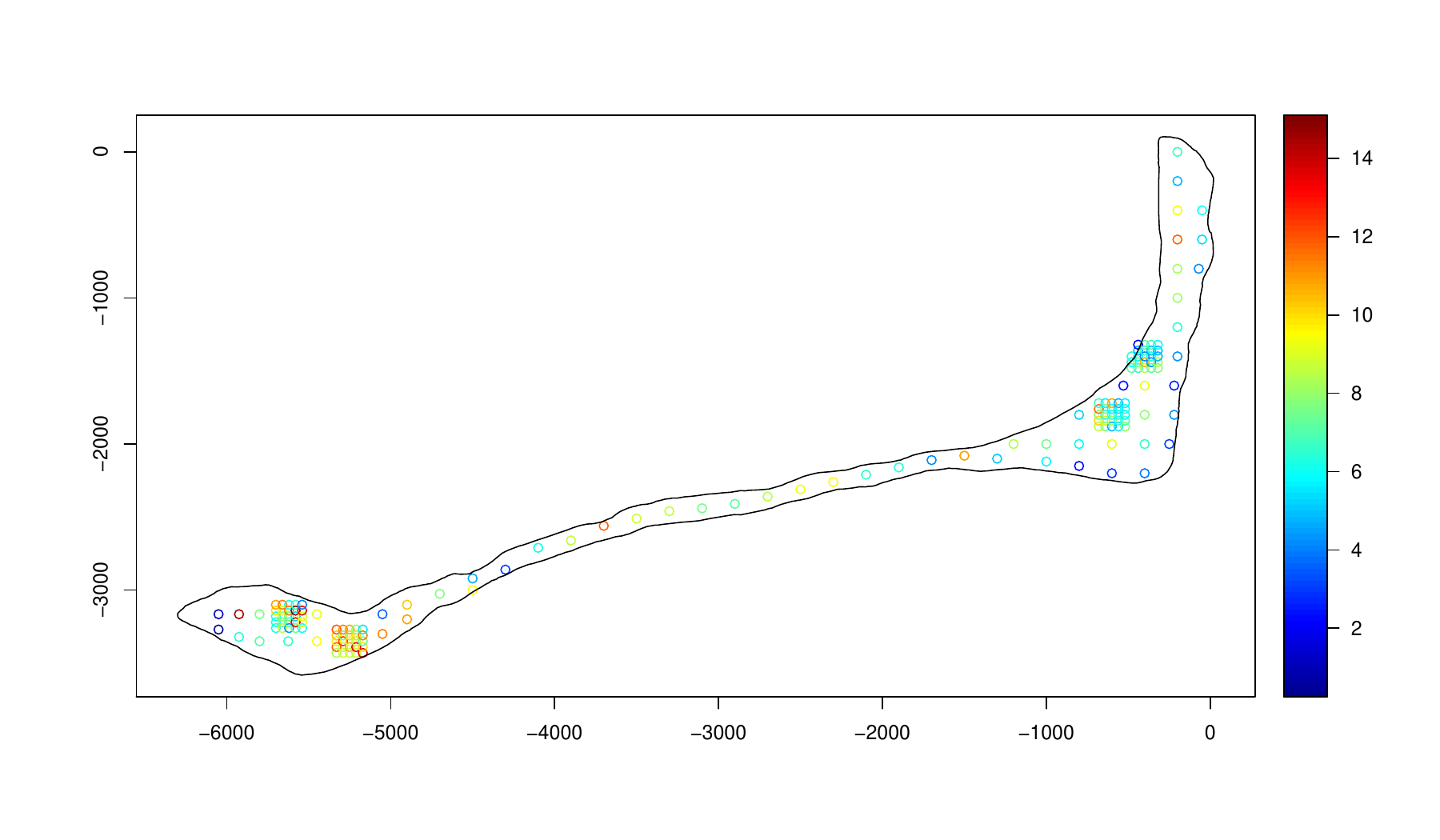}
	\end{center}
	\caption{Map of the Rongelap Island with 157 sampling locations; the different colours
		represent the $\gamma$-ray counts.}
	\label{fig:rongelap_island}
\end{figure}

As part of a wider investigation to establish whether Rongelap can safely be resettled, the Marshall Islands 
National Radiological Survey has examined the current levels of $^{137}$Cs contamination by 
\textit{in situ} $\gamma$-ray counting at a set of 157 locations over the island. Figure \ref{fig:rongelap_island}
shows the map of the Rongelap Island and the $\gamma$-ray counts at the $157$ locations.

The data consists of the following:
\begin{itemize}
	\item A $157 \times 2$ matrix which indicates the coordinates of 157 sampled locations;
	\item A vector of $\gamma$-ray counts for the 157 sampled locations;
	\item A vector of the time (in seconds) over which the 157 counts were accumulated.
\end{itemize}

Here the objective is to determine whether the island is inhabitable or not and identifying locations which still exhibit high radioactivity. In spatial models, regions are identified where a studied process exceeds a certain threshold with high probability. In most of the cases the probability of exceeding the threshold is marginally computed for each spatial location. 
We have re-framed this problem from a hypothesis testing point of view and for each of the 157 locations we test whether the intensity of radioactivity exceeds a certain threshold.

\ctn{diggle98} proposed the following model for the count data:		
\begin{align*}
Y(\bx_i)&\overset{ind}{\sim}  Poisson(t(\bx_i)\lambda(\bx_i) ),
\end{align*}
where $Y(\bx)$ is the $\gamma$-ray count, $t(\bx)$ is the time over which the counts were accumulated and $\lambda(\bx)$ is the intensity of 
radioactivity at location $\bx$, modelled as the following:
\[
\lambda(\bx) =\exp(S(\bx)),
\]		 
where $S(\bx)$ is the following \textit{Gaussian process}:
\begin{align}
&E(S(\bx))=\beta,\notag\\
&Cov(S(\bx_i),S(\bx_j))=\sigma^2\exp\left[ -\alpha\norm{\bx_i-\bx_j}^\delta \right] \label{eq:corfunc}
\end{align}
where $-\infty<\beta<\infty, \sigma>0,\alpha>0$ and $\delta>0$. 
Following \ctn{deyos16}, we set $\delta=1$, 
and propose uniform priors on $(\beta, \log(\sigma^2), \log(\alpha))$.

The hypotheses of our interest are
\begin{equation}
H_{0i}:\lambda(\bx_i) \geq c  \hbox{ vs. } H_{1i} : \lambda(\bx_i) <c,~i=1,\cdots,157,\label{eq:rong_hyp}
\end{equation}

for some appropriate threshold $c>0$.

\subsection{Multiple testing details}
\label{subsec:multiple_testing_realdata}

\subsubsection{Choices of the threshold $c$}
\label{subsubsec:threshold}
Note that based on an informal approach, \ctn{diggle98} also attempted to provide some assessment 
if the island is inhabitable. Although they did not adopt any multiple testing framework, specification
of a threshold for the intensity was still required in their case. Their specification, $c=15$, was not based
on any scientific consideration but on subjective judgement (personal communication with Peter Diggle). However,
with respect to our prior, such a threshold turned out to be too large in the sense that all the sites
turned out to be inhabitable. Rather, the $95$-th percentile of the prior of $\lambda(\bx)$ turned out 
to be close to $5$ for most locations, so that 
the choice $c=5$ seemed to be quite appropriate in our case. We also investigated with $c=10$, exceeding which
would indicate serious evidence of radioactivity in such locations. Indeed, 
the $99$-th percentile of the prior of $\lambda(\bx)$ is close to $10$ for most locations.
These choices of the threshold, ranging from $c=5$ to $c=15$ enabled us to provide some information on the 
increasing degree of severity of radionuclide concentrations in various regions of the Rongelap map.

\subsubsection{Formation of groups $G_1,\ldots,G_m$}
\label{subsubsec:groups_realdata}
In spatial analysis, locations which are physically close should exhibit similar response and high correlation. Hence, it is ideal to form groups on the basis of nearby locations. Also note that the correlation function in (\ref{eq:corfunc}) which is inversely proportional to the distances between the spatial locations. Thus forming groups on the basis of prior correlation is equivalent to forming groups of nearby locations.
. 

For each $i,~j=1,\ldots,m$, with $i<j$, we compute $\zeta_{ij}=\left(\|\bx_i-\bx_j\|\right)^{-1}$, and obtain the $95$-th percentile $\zeta$.
We then let $G_i$ to be the set consisting of those indices $j$ such that 
$\zeta_{ij}\geq \zeta$. This strategy not only is equivalent to the prior based group formation strategy in \ref{subsec:groups} but also is physically interpretation. 
It is to be seen that the group formation does not depend upon the choice of prior on the hyper-parameters. 

\subsubsection{Implementation of the Bayesian non-marginal procedure}
\label{subsubsec:model_iplementation}
To execute the multiple testing problem in (\ref{eq:rong_hyp}), the joint posterior distribution of the $\lambda(\bx_i)$s are required. The posterior distribution is approximated by drawing 
$8.5\times 10^5$ thinned samples (by storing the last one
in every 100 iterations) from the posterior distribution 
by the optimally scaled additive TMCMC method in the same way as \ctn{deyos16}.

We first test with $c=5$ to detect the locations with moderate traces of radioactivity and then 
identify the locations that show high intensity (corresponding to $c=10$). 
In each case, the estimated $\mfdr$ is less than 0.10.

\subsection{Results of multiple testing}
\label{subsubsec:results_rongelap}

\begin{figure}[h]
	\centering
	\includegraphics[trim={.55in 1in .5in 1.1in },clip, totalheight=0.225\textheight]{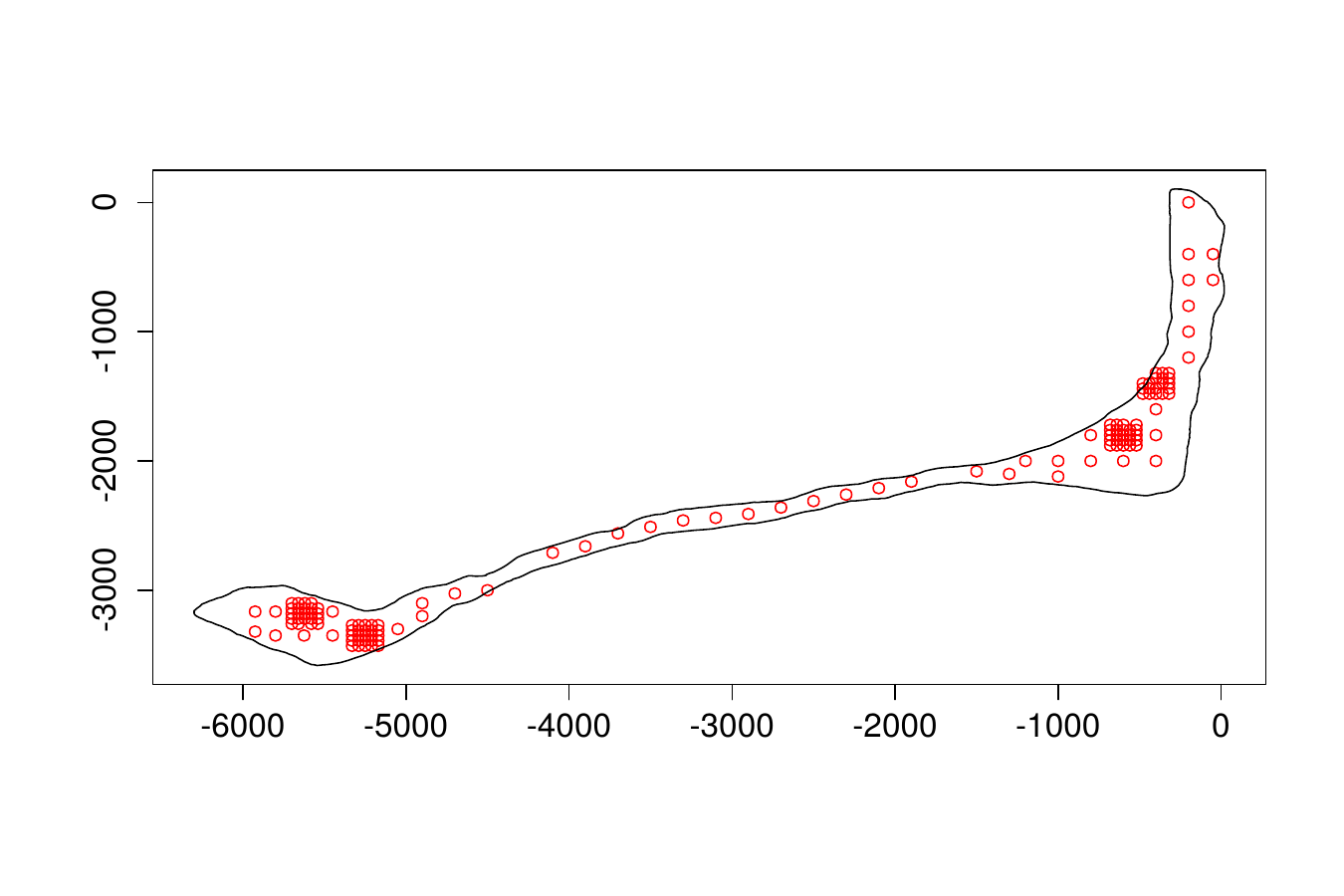}	
	\caption{The marked locations exhibit moderate traces of radioactivity (exceeding threshold $c=5$).}
	\label{fig:c5}
\end{figure}

\begin{figure}	
	\centering
	\includegraphics[trim={.55in 1in .5in 1.1in },clip, totalheight=0.225\textheight]{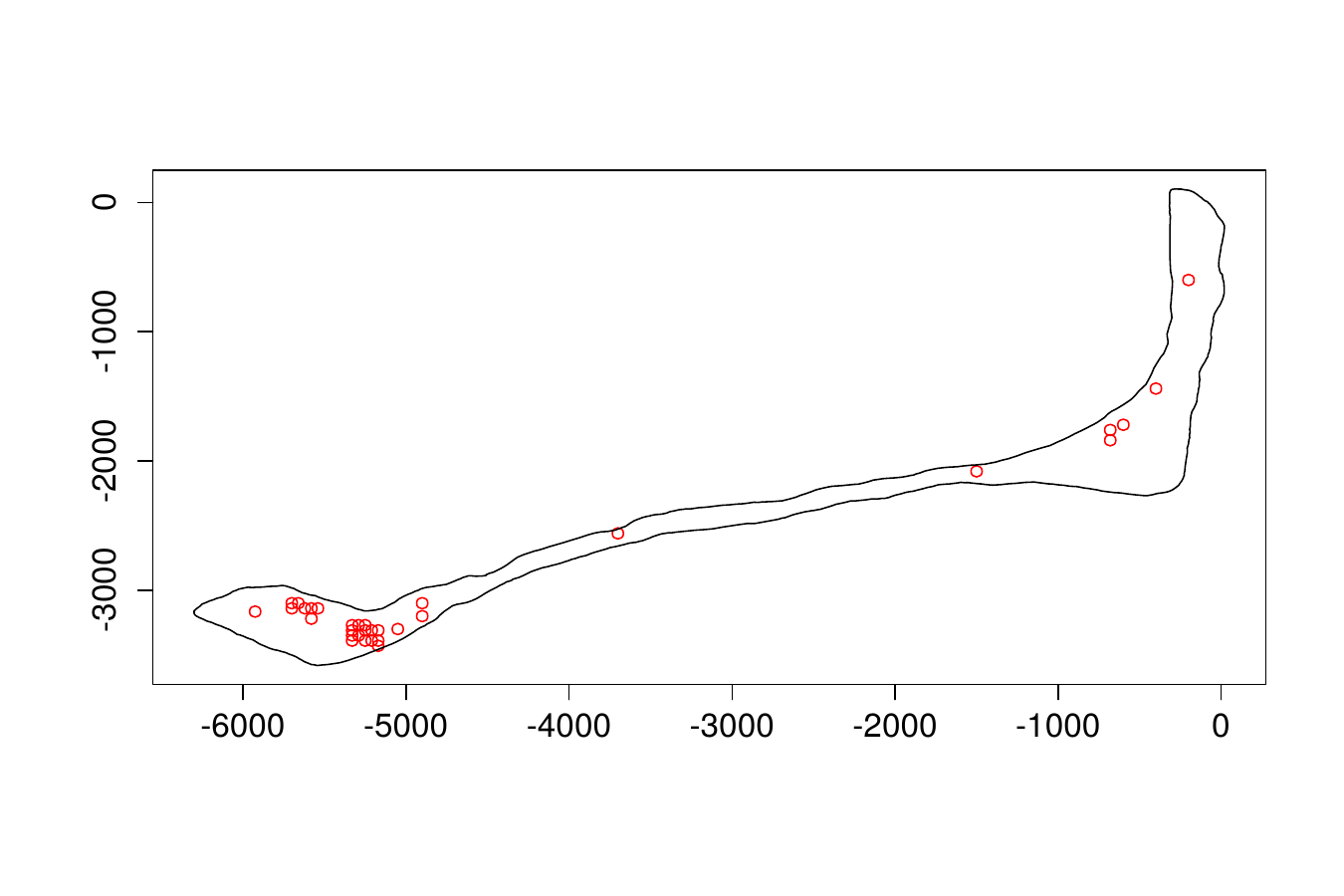}	
	\caption{The marked locations show strong signs of radioactivity (exceeding threshold $c=10$).}
	\label{fig:c10}
\end{figure}

The locations marked in Figure \ref{fig:c5} show moderate traces of radioactivity and
those of Figure \ref{fig:c10} have high intensity of radioactivity and are not inhabitable. On the basis of the data, many locations are exhibiting traces of radioactivity, especially Figure \ref{fig:c10} showing several locations with high radioactivity. 

\section{Summary and conclusion}
\label{sec:conclusion}
In this work we have proposed and developed a novel and general Bayesian multiple testing procedure that yields joint decisions
regarding the dependent hypotheses, via the relevant joint posterior probabilities. In keeping with the concept,
we have proposed a new Bayesian version of $pFDR$, namely, $mpBFDR$, which appropriately takes the dependent 
situation into account, and possesses desirable theoretical properties. Interestingly, our simulation study showed that 
in dependent situations, even for established marginal multiple testing methods, it makes more sense to control 
$mpBFDR$ rather than $pBFDR$, in order to have a higher chance of capturing the true decision configuration. \ctn{Chandra16} showed strong consistency of the non-marginal procedure under general dependence.
In another simulation study in the dependent scenario, our non-marginal procedure significantly outperformed 
the popular existing marginal methods in terms of lesser $pBFNR$. Application of our ideas to a real spatial
data set also yielded encouraging results.  

Indeed, in most practical applications of multiple testing problem, testing thousands of hypotheses in isolation 
is seldom meaningful. In popular applications like detecting  bio-markers from microRNA expression data or 
in neuroscience, where millions of parameters are of interest, 
the parameters bound to have strong dependence structure among themselves. If the underlying multiple testing
procedure pays less attention to the dependence structure, giving importance mostly to multiplicity adjustment 
with marginal p-values, Bayes factors, or marginal 
posterior probabilities, then it may miss insightful information, leading to less interesting results. 
As we demonstrated, our proposed procedure appropriately balances both the issues, which is instrumental in
significantly outperforming the existing, established multiple testing methods.

However, our method heavily depends on suitable selection of dependent groups for each hypothesis. In the simulation study
reported in Section \ref{sec:simulation_study}, 
the exact dependence structure between the parameters is known. 
In the real data analysis discussed in Section \ref{sec:realdata}, the group formation is straightforward since 
the covariance is a decreasing function of the geographical distances. In other practical problems, group selection 
may not be as simple. In such cases, some data driven procedure needs to be adopted. Forming suitable groups not only 
improves the inference in multiple testing, but also helps understand the joint behaviour of the concerned parameters 
which often may be the subject of interest. This paper gives rise to these interesting but challenging problems to 
venture for our future work.


\newpage
\begin{center}
	{\LARGE\bf Supplementary Material}
\end{center}

\renewcommand\thefigure{S-\arabic{figure}}
\renewcommand\thetable{S-\arabic{table}}
\renewcommand\thesection{S-\arabic{section}}
\renewcommand{\theequation}{S-\arabic{equation}}

\section{ Break-up of the number of hypotheses being tested into error and non-error terms}
\label{sec:appendix2}
We denote the error terms by $E$ and the non-error terms by $NE$.
\begin{itemize}
	\item[(1)] $NE_1=\sum_{i=1}^md_ir_iz_i$, equalling $\#\{i:d_i=1,r_i=1,z_i=1\}$. In words, the term
	corresponds to the number of cases where $H_{1i}$ are correctly accepted, and all other decisions are
	also correct.
	\item[(2)] $NE_2=\sum_{i=1}^m(1-d_i)(1-r_i)z_i$, equalling $\#\{i:d_i=0,r_i=0,z_i=1\}$. In words, the term
	corresponds to the number of cases where $H_{1i}$ are correctly rejected, and all the remaining decisions are
	correct.
	\item[(3)] $E_1=\sum_{i=1}^md_i(1-r_i)z_i$, equalling $\#\{i:d_i=1,r_i=0,z_i=1\}$. In words, the term
	corresponds to the number of cases where $H_{1i}$ are wrongly accepted, but all the remaining decisions are
	correct.
	\item[(4)] $E_2=\sum_{i=1}^md_i(1-r_i)(1-z_i)$, equalling $\#\{i:d_i=1,r_i=0,z_i=0\}$. In words, the term
	corresponds to the number of cases where $H_{1i}$ are wrongly accepted, but at least one of the remaining decisions is
	incorrect.
	\item[(5)] $E_3=\sum_{i=1}^md_ir_i(1-z_i)$, equalling $\#\{i:d_i=1,r_i=1,z_i=0\}$. In words, the term
	corresponds to the number of cases where $H_{1i}$ are correctly accepted, and at least one of the remaining decisions 
	is incorrect.
	\item[(6)] $E_4=\sum_{i=1}^m(1-d_i)(1-r_i)(1-z_i)$, equalling $\#\{i:d_i=0,r_i=0,z_i=0\}$. In words, the term
	corresponds to the number of cases where $H_{1i}$ are correctly rejected, but at least one the remaining decisions is
	incorrect.
	\item[(7)] $E_5=\sum_{i=1}^m(1-d_i)r_iz_i$, equalling $\#\{i:d_i=0,r_i=1,z_i=1\}$. In words, the term
	corresponds to the number of cases where $H_{1i}$ are wrongly rejected, but all the remaining decisions are
	correct.
	\item[(8)] $E_6=\sum_{i=1}^m(1-d_i)r_i(1-z_i)$, equalling $\#\{i:d_i=0,r_i=1,z_i=0\}$. In words, the term
	corresponds to the number of cases where $H_{1i}$ are wrongly rejected, and at least one of the remaining decisions is
	incorrect.
	
\end{itemize}
Clearly, $$NE_1+NE_2+E_1+E_2+E_3+E_4+E_5+E_6=m.$$

\subsection{Proof of Theorem \ref{th:overpenalty}}
\label{subsec:proof_overpenalty}
We first recall that (see Section \ref{sec:err} of the main manuscript) our idea is to maximize $g_1(\bd)=TP(\bd)-\lambda E(\bd)$, 
where $E(\bd)=E_1(\bd)+E_2(\bd)+E_3(\bd)$, with respect to the decision configuration $\bd$, where $\lambda>0$. Let us consider another function
$g_2(\bd)=TP(\bd)-\lambda \left(E(\bd)+E^*(\bd)\right)$, where $E^*(\bd)$ is another error term. Let $\hat\bd_1$ and $\hat\bd_2$ denote the maximizers of $g_1$ and $g_2$, respectively.
Let $\beta=\frac{\lambda}{1+\lambda}$. Now, there exists $c~(>0)$ such that $E^*(\hat\bd_2)=cE(\hat\bd_2)$. Then $\lambda \left(E(\hat\bd_2)+E^*(\hat\bd_2)\right)=\left(\lambda+c\right)E(\hat\bd_2)$.
Let $\beta^*=\frac{\lambda+c}{1+\lambda+c}$.
The remaining part of the proof follows similarly as the proof of Lemma \ref{lemma:discovery} 
of our main manuscript, but here we present the details for clarity. Letting $f_1(\bd)=\sum d_i\left(w_i(\bd)-\beta\right)$ it follows from the definition of maximization that
\begin{align}
&g_2(\hat\bd_2)\geq g_2(\hat\bd_1)\notag\\
\Rightarrow &\sum\hat d_{2i}\left(w_i(\hat\bd_2)-\beta^*\right)\geq \sum\hat d_{1i}\left(w_i(\hat\bd_1)-\beta^*\right)\notag\\
\Rightarrow &f_1(\hat \bd_2)-(\beta^*-\beta)\sum \hat d_{2i}\geq f_1(\hat \bd_1)-(\beta^*-\beta)\sum \hat d_{1i}\notag\\
\Rightarrow &f_1(\hat\bd_2)- f_1(\hat\bd_1)\geq \left(\beta^*-\beta\right)\left(\sum \hat d_{2i}-\sum \hat d_{1i}\right).
\label{eq:eq1}
\end{align}
Now $\beta^*-\beta=\frac{c}{(1+\lambda)(1+\lambda+c)}>0$. 
Hence, we must have $\sum \hat d_{2i}\leq \sum \hat d_{1i}$, otherwise
$f_1(\hat\bd_2)- f_1(\hat\bd_1)>0$, which would contradict the fact that $\hat\bd_1$ is the maximizer associated with $ f_1$. In other words, the number of rejections
in the decision configuration $\hat\bd_2$, the maximizer associated with the extra error term $E^*$, is less than or equal to that in $\hat\bd_1$, the maximizer corresponding
to the procedure with less error terms to be controlled. Thus, controlling many error terms would lead to false acceptance of most alternative hypotheses.

\section{Assumptions of \ctn{Shalizi09}}
\label{subsec:assumptions_shalizi}
\begin{enumerate}[label={(S\arabic*)}]	
	\item  \label{shalizi1} Consider the following likelihood ratio:
	\begin{equation}
	R_n(\btheta)=\frac{f_{\btheta}(\bX_n)}{p(\bX_n)}.
	\label{eq:R_n}
	\end{equation}
	Assume that $R_n(\btheta)$ is $\sigma(\bX_n)\times \mathcal T$-measurable for all $n>0$.
	
	\item \label{s2} For each $\btheta\in\Theta$, the generalized or relative asymptotic equipartition property holds, and so,
	almost surely,
	\begin{equation*}
	\underset{n\rightarrow\infty}{\lim}~\frac{1}{n}\log R_n(\btheta)=-h(\btheta),
	\end{equation*}
	where $h(\btheta)$ is given in (S3) below.
	
	\item \label{s3} For every $\btheta\in\Theta$, the KL-divergence rate
	\begin{equation}
	h(\btheta)=\underset{n\rightarrow\infty}{\lim}~\frac{1}{n}E\left(\log\frac{p(\bX_n)}{f_{\btheta}(\bX_n)}\right).
	\label{eq:S3}
	\end{equation}
	exists (possibly being infinite) and is $\mathcal T$-measurable.
	
	\item \label{s4}
	Let $I=\left\{\btheta:h(\btheta)=\infty\right\}$. 
	The prior $\pi$ satisfies $\pi(I)<1$. 
	
	\item \label{s5} There exists a sequence of sets $\mathcal G_n\rightarrow\Theta$ as $n\rightarrow\infty$ 
	such that: 
	\begin{enumerate}
		\item[(1)]
		\begin{equation}
		\pi\left(\mathcal G_n\right)\geq 1-\alpha\exp\left(-\varsigma n\right),~\mbox{for some}~\alpha>0,~\varsigma>2h(\Theta);
		\label{eq:S5_1}
		\end{equation}
		\item[(2)]The convergence in (S3) is uniform in $\theta$ over $\mathcal G_n\setminus I$.
		\item[(3)] $h\left(\mathcal G_n\right)\rightarrow h\left(\Theta\right)$, as $n\rightarrow\infty$.
	\end{enumerate}
	
	For each measurable $A\subseteq\Theta$, for every $\delta>0$, there exists a random natural number $\tau(A,\delta)$
	such that
	\begin{equation}
	n^{-1}\log\int_{A}R_n(\btheta)\pi(\btheta)d\btheta
	\leq \delta+\underset{n\rightarrow\infty}{\limsup}~n^{-1}
	\log\int_{A}R_n(\btheta)\pi(\btheta)d\btheta,
	\label{eq:limsup_2}
	\end{equation}
	for all $n>\tau(A,\delta)$, provided 
	$\underset{n\rightarrow\infty}{\lim\sup}~n^{-1}\log\pi\left(\mathbb I_A R_n\right)<\infty$.
	Regarding this, the following assumption has been made by Shalizi:
	
	\item\label{s6} The sets $\mathcal G_n$ of (S5) can be chosen such that for every $\delta>0$, the inequality
	$n>\tau(\mathcal G_n,\delta)$ holds almost surely for all sufficiently large $n$.
	
	\item \label{shalizi7} The sets $\mathcal G_n$ of (S5) and (S6) can be chosen such that for any set $A$ with $\pi(A)>0$, 
	\begin{equation}
	h\left(\mathcal G_n\cap A\right)\rightarrow h\left(A\right)\text{ as } n\rightarrow\infty. 
	\label{eq:S7}
	\end{equation}	
\end{enumerate}

\section{Proofs of Theorems and Lemmas}
\subsection{Proof of Theorem \ref{theorem:theorem1}}	
For our purpose, we first state and prove a lemma.

\begin{lemma}
	\label{lemma:lemma1}
	For $\ell=1,\ldots,k$, let, for $q\geq 1$, $g_\ell:\Rn{q}\rightarrow\Rn{}$ be a continuous function, with
	$\abs{g_\ell}<M<\infty$. 
	Consider a sequence 
	$\{\beta_b\}_{b=1}^{\infty}$ converging to $\beta\in\Rn{}$. 
	Define 
	$$A_b= \bigcap_{\ell=1}^k g_\ell^{-1} (a_\ell\beta_b,M) \ \ and \ \ A= \bigcap_{\ell=1}^k g_\ell^{-1} (a_\ell\beta,M),$$
	where, for $\ell=1,\ldots,k$, $a_\ell\in\Rn{}$.
	Consider any measure $\mu$ satisfying 
	\begin{equation}
	\mu\left[  \bigcap_{\ell=1}^k\{x: g_\ell(x)=a_\ell\beta\}\right] =0. 
	\label{eq:mu_zero}
	\end{equation}
	Then, for any bounded function $h:\Rn{q}\rightarrow\Rn{}$ integrable with respect to $\mu$,
	it holds that
	$$\underset{b\rightarrow\infty}{\lim}\int h(x) I_{A_b}(x)d\mu(x)=\int h(x) I_A(x)d\mu(x).$$
\end{lemma}
\begin{proof}
	First note that,
	\begin{align*}
	\underset{b\rightarrow\infty}{ \lim\inf}~I_{A_b}(x)&=I_{\lim\inf A_b} (x);\\
	\underset{b\rightarrow\infty}{ \lim\sup}~I_{A_b}(x)&=I_{\lim\sup A_b} (x).
	\end{align*}
	Consider $x\in A$. Then $g_\ell(x)>a_\ell\beta,\ \forall\ \ell=1,2,\ldots ,k$. 
	Take $\epsilon<\underset{\ell=1,\ldots,k}{\min}(g_\ell(x)-a_\ell\beta )$.\\ Then 
	\begin{align}
	&\exists\ b_0\in\mathbb{N}\ \ni\ \mbox{for}~b>b_0, \ a_\ell\beta_b<a_\ell\beta
	+\epsilon<g_\ell(x),\ \forall\ \ell=1,2,\cdots ,k;\\\nonumber
	\Rightarrow~	& x\in A_b\ \forall\ b>b_0 \Rightarrow x\in \lim\inf A_b\\\nonumber
	\Rightarrow~	& A\subseteq \lim\inf A_b.
	\end{align}
	Now take
	\begin{align}
	&x\in\lim\inf A_b \Rightarrow \exists\ b_1\in\mathbb{N}\ \ni \ x\in A_b\ \forall \ b>b_1;\\\nonumber
	\Rightarrow~ & g_\ell(x)>a_\ell\beta_b\ \forall \ b>b_1 \Rightarrow g_\ell(x)\geq a_\ell\beta\ \forall\ \ell=1,2,\cdots,k;\\
	\nonumber
	\Rightarrow~ & x\in \bigcap_{\ell=1}^k g_\ell^{-1}[a_\ell\beta,M).
	\end{align}
	
	Next consider $x\in\lim\sup A_b\setminus A$. Then
	$x\in\lim\sup A_b\Rightarrow\exists$ a subsequence $\{b_j\}_{j=1}^{\infty} \ \ni \ x\in A_{b_j}$, for $j=1,2,\ldots$.
	That is, for $\ell=1,2,\ldots,k$, and for $j=1,2,\ldots$, $g_\ell(x)>a_\ell\beta_{b_j}\Rightarrow g_\ell(x)\geq a_\ell\beta\ 
	\forall\ \ell=1,2,\ldots ,k$.
	Again, $x\in A^c\Rightarrow g_\ell(x)\leq a_\ell\beta$, for $\ell=1,2,\ldots ,k$.
	Hence,  $ g_\ell(x)=a_\ell\beta\ \forall\ \ell=1,2,\cdots ,k\Rightarrow x\in \bigcap_{\ell=1}^k\{x: g_\ell(x)=a_\ell\beta\}$.
	
	It follows that
	$$A\subseteq\lim\inf A_b\subseteq\lim\sup A_b\subseteq\bigcap_{\ell=1}^k g_\ell^{-1}[a_\ell\beta,M).$$
	Now, let
	\begin{align*}
	\lim\inf I_{A_b}(x)= I_{\lim\inf A_b} (x)= I_A+I_L;\\
	\lim\sup I_{A_b}(x)= I_{\lim\sup A_b} (x)= I_A+I_U,
	\end{align*}
	where $$L\subseteq U\subseteq \bigcap_{\ell=1}^k\{x: g_\ell(x)=a_\ell\beta\}.$$
	Using (\ref{eq:mu_zero}) it is easily seen that
	$$\lim\inf I_{A_b}(x)=\lim\sup I_{A_b}(x)=I_A,\ \ \mu\mbox{-almost everywhere}.$$
	It follows from the above that $h(x)I_{A_b}(x)$ is a bounded function converging point wise to $h(x)I_A(x)$, 
	$\mu$-almost everywhere. Hence, using the dominated convergence theorem we conclude that
	$$\underset{b\rightarrow\infty}{\lim}\int h(x) I_{A_b}(x)d\mu(x)=\int h(x) I_A(x)d\mu(x).$$
\end{proof}

In our case,
\begin{align*}
mpBFDR =& E_{\bX_n} \left[\sum_{\dec\in\mathbb{D}} \frac{\sum_{i=1}^{m}d_i(1- w_i(\dec) )}{\sum_{i=1}^{m}d_i}
\delta_\beta(\dec|\bX_n)\bigg{|}\delta_\beta(\dec=\bzero|\bX_n)=0 \right]\\
=&\sum_{\dec\in\mathbb{D}} E_{\bX_n} \left[\frac{\sum_{i=1}^{m}d_i(1- w_i(\dec) )}{\sum_{i=1}^{m}d_i}
\delta_\beta(\dec|\bX_n)\bigg{|}\delta_\beta(\dec=\bzero|\bX_n)=0 \right]\\
=&\sum_{\dec\in\mathbb{D}} E_{\bX_n} \left[\frac{\sum_{i=1}^{m}d_i(1- w_i(\dec) )}{\sum_{i=1}^{m}d_i} 
I\left( \sum_{i=1}^{m}d_i>0 \right)\delta_\beta(\dec|\bX_n) \right] 
\frac{1}{P_{\bX_n}\left[ \delta_\beta(\dec=\bzero|\bX_n)=0\right] } \\
=&\sum_{\dec\in\mathbb{D}\setminus\left\lbrace \bzero\right\rbrace}
E_{\bX_n}\left[\frac{\sum_{i=1}^{m}d_i(1- w_i(\dec) )}{\sum_{i=1}^{m}d_i} 
\delta_\beta(\dec|\bX_n) \right] \frac{1}{P_{\bX_n}\left[ \delta_\beta(\dec=\bzero|\bX_n)=0\right] }, 
\end{align*}
where $\beta\in (0,1)$, rather than $\beta\in\mathbb R$ used in Lemma \ref{lemma:lemma1} for greater generality.

To prove continuity of $mpBDFR$ with respect to $\beta$ it is enough to show that 
\\
$E_{\bX_n}\left[\frac{\sum_{i=1}^{m}d_i(1- w_i(\dec) )}{\sum_{i=1}^{m}d_i} 
\delta_\beta(\dec|\bX_n) \right]$ and $\frac{1}{P_{\bX_n}\left[ \delta_\beta(\dec=\bzero|\bX_n)=0\right]}$ 
are continuous with respect to $\beta$ for all $\dec\neq \bzero$.

To prove continuity of 
$E_{\bX_n} \left[\frac{\sum_{i=1}^{m}d_i(1- w_i(\dec) )}{\sum_{i=1}^{m}d_i} \delta_\beta(\dec|\bX_n) \right]$
observe that $\delta_{\beta}(\dec|\bX_n)$ is the indicator of the set
$$\bigcap_{\dec^\ast\neq\dec}
\left\{\bX_n:\sum_{i=1}^m d_iw_i(\dec)-\sum_{i=1}^m d^\ast_iw_i(\dec^\ast)
>\beta\left(\sum_{i=1}^md_i-\sum_{i=1}^md^\ast_i \right)\right\},$$
so that referring to Lemma \ref{lemma:lemma1} we identify
$g_{\ell}\equiv\sum_{i=1}^m d_iw_i(\dec)-\sum_{i=1}^m d^\ast_iw_i(\dec^\ast)$; $\ell=1,\ldots,k$, where
$k=2^m-1$ (the number of decision configurations except $\bd$), 
where $\ell$ indexes $\dec^\ast$. Also note that $a_\ell=\sum_{i=1}^md_i-\sum_{i=1}^md^*_i$
and that $h\equiv\frac{\sum_{i=1}^{m}d_i(1- w_i(\dec) )}{\sum_{i=1}^{m}d_i}$, which is a bounded function.
The assumption that the event $\left\{\bX_n:g_{\ell}(\bX_n)=a_\ell\beta\right\}$ has zero probability,
in conjunction with Lemma \ref{lemma:lemma1}, then lets us conclude that 
$E_{\bX_n} \left[\frac{\sum_{i=1}^{m}d_i(1- w_i(\dec) )}{\sum_{i=1}^{m}d_i} \delta_\beta(\dec|\bX_n) \right]$
is continuous with respect to $\beta$. 

To see continuity of $P_{\bX_n}\left[ \delta_\beta(\dec=\bzero|\bX_n)=0\right]$ with respect to $\beta$,
note that this probability is the same as 
$
1-P_{\bX_n}\left(-\sum_{i=1}^m d^\ast_iw_i(\dec^\ast)>-\beta\sum_{i=1}^md^\ast_i;~\forall~\dec^\ast\neq\bzero \right),
$
from which we can easily identify, referring to Lemma \ref{lemma:lemma1}, that 
$g_{\ell}\equiv-\sum_{i=1}^m d^\ast_iw_i(\dec^\ast)$, $a_\ell=-\sum_{i=1}^md^*_i$, and $h\equiv 1$,
so that Lemma \ref{lemma:lemma1} also guarantees continuity of $P_{\bX_n}\left[ \delta_\beta(\dec=\bzero|\bX_n)=0\right]$
with respect to $\beta$.

Hence, Theorem \ref{theorem:theorem1} is proved.

\subsection{Proof of Lemma \ref{lemma:discovery}}
Let $\dec'=\underset{\dec\in\mathbb{D}} {\argmax} f_{\beta'}(\dec)$ and 
$\dec''=\underset{\dec\in\mathbb{D}} {\argmax} f_{\beta''}(\dec)$ where $\beta''>\beta'$. 
Note that,
\begin{align}
& f_{\beta''}(\dec'')\geq f_{\beta''}(\dec')\nonumber\\
\Rightarrow~ &\sum d''_i w_i(\dec'')-\beta'\sum d''_i -(\beta''-\beta')\sum d''_i
\geq \sum d'_i w_i(\dec')-\beta'\sum d'_i -(\beta''-\beta')\sum d'_i;\nonumber\\
\Rightarrow~ &f_{\beta'}(\dec'')-f_{\beta'}(\dec')\geq(\beta''-\beta')\sum (d''_i-d'_i). \label{eq:ddn}
\end{align}
If $\sum d'_i<\sum d''_i$, then the right hand side of (\ref{eq:ddn}) will be greater than 0, 
contradicting the fact that $\dec'=\underset{\dec\in\mathbb{D}} {\argmax} f_{\beta'}(\dec)$. Hence, $\sum d'_i\geq\sum d''_i$.	

\subsection{Proof of Theorem \ref{theorem:theorem2}}

As in the proof of Lemma \ref{lemma:discovery}, let $\dec'=\underset{\dec\in\mathbb{D}} {\argmax} f_{\beta'}(\dec)$ and 
$\dec''=\underset{\dec\in\mathbb{D}} {\argmax} f_{\beta''}(\dec)$ where $\beta''>\beta'$.  
If possible, let
\begin{align}
&\frac{\sum d''_i (1- w_i(\dec''))}{\sum d''_i} > \frac{\sum d'_i (1-w_i(\dec'))}{\sum d'_i}\label{eq:assump1};\\
\Rightarrow &\sum d''_i \sum d'_i w_i(\dec') >\sum d'_i \sum d''_i  w_i(\dec''). \label{eq:dcr}
\end{align}
Again,
\begin{align}
f_{\beta''}(\dec'')&=\frac{1}{\sum d'_i}\left\lbrace \sum d'_i\sum d''_i  w_i(\dec'')- \sum d'_i\sum d''_i \beta''\right\rbrace\\
& < \frac{1}{\sum d'_i}\left\lbrace \sum d''_i \sum d'_i w_i(\dec')- \sum d'_i\sum d''_i \beta''\right\rbrace 
\text{[from (\ref{eq:dcr})]}\\
&= \frac{\sum d''_i}{\sum d'_i}f_{\beta''}(\dec')\\
&\leq f_{\beta''}(\dec').\label{eq: contra}
\end{align}
Thus, (\ref{eq: contra}) contradicts the fact that $\dec''=\underset{\dec\in\mathbb{D}} {\argmax} f_{\beta''}(\dec)$. 
Hence, (\ref{eq:assump1}) is not possible and we have that
\begin{align}
\frac{\sum d''_i (1- w_i(\dec''))}{\sum d''_i} < \frac{\sum d'_i (1-w_i(\dec'))}{\sum d'_i}. \label{eq:decineq}
\end{align} 

Note that, for any $0<\beta<1$, $P_{\bX_n}[\delta_\beta(\dec'|\bX_n)=0] =P_{\bX_n}[f_\beta(\dec')<f_\beta(\dec)\text{ for at least one } \dec\neq\dec']$.
\begin{align*}
\therefore  P_{\bX_n}[\delta_{\beta}(\dec=\bzero|\bX_n)=0]&=P_{\bX_n}[f_{\beta}(\dec)>0\text{ for at least one } \dec\neq\bzero]\\
&=P_{\bX_n}\left[ \sum d_i w_i(\dec)>\beta\sum d_i \text{ for at least one } \dec\neq\bzero\right]
\end{align*}
Define, $A_\beta=\left\lbrace \sum d_i w_i(\dec)>\beta\sum d_i \text{ for at least one } \dec\neq\bzero\right\rbrace$. 
Clearly, $A_{\beta''}\subset A_{\beta'}$ for $\beta''>\beta'$. Hence,
\begin{align}
E_{\bX_n}\left[ \frac{\sum d'_i (1-w_i(\dec'))}{\sum d'_i}\bigg{|} A_{\beta'} \right]
&=\int_{A_{\beta'}} \frac{\sum d'_i (1-w_i(\dec'))}{\sum d'_i} dP_{\bX_n}\\
&\geq \int_{A_{\beta''}} \frac{\sum d'_i (1-w_i(\dec'))}{\sum d'_i} dP_{\bX_n}\\
&\geq \int_{A_{\beta''}} \frac{\sum d''_i (1-w_i(\dec''))}{\sum d''_i} dP_{\bX_n}~~\mbox{[by (\ref{eq:decineq})]}\\
&=E_{\bX_n}\left[\frac{\sum d''_i (1-w_i(\dec''))}{\sum d''_i}\bigg{|} A_{\beta''} \right].\label{eq:mbfdrineq}
\end{align}
Since, $\dec''$ and $\dec'$ are the maximizers of $f_{\beta''}(\dec)$ and $f_{\beta'}(\dec)$, 
the left hand and right hand sides of (\ref{eq:mbfdrineq}) boil down to $mpBFDR$s with respect to 
the penalization constants $\beta''$ and $\beta'$ respectively, associated with the non-marginal method. 
This proves the theorem.

\subsection{Proof of Theorem \ref{theorem:compare_methods}}
For a decision configuration $\bd$, we define the following sets:
\begin{align*}
\mathcal{I}(\bd)&=\left\{i:d_i=1\right\}\\
\mathcal{I}(\bd)^c&=\{1,\cdots,m\}\backslash \mathcal{I}(\bd).
\end{align*}
Note that $\mathcal{I}(\bd)$ is the set of hypotheses where the null hypotheses are rejected. Now, for $i\in\mathcal{I}(\bd)$
\begin{align*}
w_i(\bd)=\postp \left(H_{1i}\cap \left\lbrace\cap_{j\neq i}H_{d_jj} \right\rbrace  \right)=\postp \left(\left\lbrace\cap_{j\in \mathcal{I}(\bd)}H_{1j} \right\rbrace\cap \left\lbrace\cap_{j\in \mathcal{I}(\bd)^c}H_{0j} \right\rbrace  \right),
\end{align*}
that is, $w_i(\bd)=w_k(\bd)$, for all $k\in\mathcal{I}(\bd)$. Therefore, we omit the suffix and write $w_i(\bd)$ as $w(\bd)$.
We state this below in the form of a lemma.
\begin{lemma}
	\label{lemma:equal_posteriors}
	Note that for any decision configuration $\bd\neq\bzero$, $\mathcal I(\bd)$ is a non-empty set. Then $w_i(\bd)=w(\bd)$ for all $i\in\mathcal I(\bd)$.
\end{lemma}
Clearly from Lemma \ref{lemma:equal_posteriors} it follows that 
$\sum_{i=1}^m\hat d_i\left(w_i(\hat\bd)-\beta\right)
=\left(\sum_{i=1}^m\hat d_i\right)\left(w(\hat\bd)-\beta\right)$.
Now, Lemma \ref{lemma:discovery} shows that $\sum_{i=1}^m\hat d_i$ is decreasing in $\beta$ for our non-marginal multiple testing procedure. Hence, for any other decision configuration $\bd^*$ corresponding to any other multiple testing method, there exists $\hat\beta$ such that $\sum_{i=1}^m\hat d_i=\sum_{i=1}^md^*_i$ for small sample size $n\geq 1$,
and hence for $\beta=\hat\beta$, $\hat\bd$ is better than $\bd^*$ in the sense of maximizing the posterior
$w(\bd)=\postp\left(\cap_{i=1}^mH_{d_{i},i}\right)$ with respect to all possible decision configurations
subject to $\sum_{i=1}^m d_{i}=\sum_{i=1}^md^*_i$, which is equivalent to minimization of the 
posterior expected ``0-1" loss subject to $\sum_{i=1}^m d_{i}=\sum_{i=1}^md^*_i$. Hence, the proof.

\subsection{Proof of Theorem \ref{theorem:compare_methods2}}
Note that, $o_r>1$ implies that more than one $\bd$ would yield same $\tilde{\bk}(\bd)$ vector. Therefore, we define the following sets:
\begin{align*}
&\mathbb D_{\tilde\bk}=\left\{\bd:k_{r}(\bd_{G^*_r})=\tilde k_{r}~\forall~r=1,\ldots,s\right\};\\
&\mathbb D_{\bS(k)}=\bigcup_{\tilde\bk\in\bS(k)}\mathbb D_{\tilde\bk}.
\end{align*}
Note that $\mathbb D_{\bS(k)}$ is the set of all decision configurations with $k$ discoveries. Let
\begin{equation}
\tilde{\bd}=\argmin_{\bd\in\mathbb D_{\bS(k)}}\pexp L(\bd^t,\bd|k)
\end{equation} 
be the optimal decision configuration corresponding to the loss function $L(\bd^t,\bd|k)$.

We define $w_{r}(\bd_{G^*_r})=\postp\left(\cap_{j\in G^*_r}H_{d_jj} \right)$. Then 
for all $i\in G_r^*$ such that $d_i=1$, we have from Lemma \ref{lemma:equal_posteriors} $w_i(\bd)=w_{r}(\bd_{G^*_r})$. Hence it follows that 
\begin{align}
&\pexp\left( L(\bd^t,\bd|k)\right)=\min_{\bd\in\mathbb D_{\bS(k)}} \sum_{r=1}^s k_r(\bd) (1- w_r(\bd_{G^*_r}))=k- \max_{\bd\in\mathbb D_{\bS(k)}} \sum_{r=1}^sk_r(\bd)  w_r(\bd_{G^*_r})\\
\Rightarrow~& \tilde{\bd}=\argmax_{\bd\in\mathbb D_{\bS(k)}}\sum_{r=1}^s k_r(\bd)  w_r(\bd_{G^*_r}).\label{eq:dtilde}
\end{align}	 
Also for the non-marginal method  we see that
$$f_\beta(\bd)=\sum_{i=1}^m d_i\left(w_i(\bd)-\beta\right)
=\sum_{r=1}^s\left(\sum_{i\in G^*_r} d_{i}\right)\left(w_{r}(\bd_{G^*_r})-\beta\right)
=\sum_{r=1}^sk_{r}(\bd)\left(w_{r}(\bd_{G^*_r})-\beta\right).$$ 
Now subject to the restriction $\sum_{i=1}^m\hat d_i=k$, from Definition \ref{def:nmd} we have	 
\begin{align}
\hat{\bd}=\argmax_{\bd\in\mathbb D_{\bS(k)}} f_\beta(\bd)=\argmax_{\bd\in\mathbb D_{\bS(k)}} \sum_{r=1}^sk_{r}(\bd)\left(w_{r}(\bd_{G^*_r})-\beta\right)=\argmax_{\bd\in\mathbb D_{\bS(k)}} \sum_{r=1}^sk_{r}(\bd) w_{r}(\bd_{G^*_r}).\label{eq:dhat}
\end{align}	 
Hence, from \eqref{eq:dtilde} and \eqref{eq:dhat} we see that the non-marginal procedure has the desirable decision theoretic property.

\subsection{Proof of Theorem \ref{th:single_group}}
Following the proof of the Theorem \ref{theorem:compare_methods} we see that the non-marginal procedure maximizes the posterior probability $w(\bd)=\postp\left(\cap_{i=1}^mH_{d_{i},i}\right)$ with respect to all possible decision configurations
subject to $\sum_{i=1}^m d_{i}=k$. Now from Theorem \ref{th:shalizi}, we have $\frac{1}{n}\log w(\bd)\approx -J(\Theta_{\bd})$ for sufficiently large $n$. 
Hence the KL-divergence rate $J(\Theta_{\bd})$ is minimized.

\subsection{Proof of Theorem \ref{th:multiple_group}}
Note that the maximization problem is equivalent to maximization of
$p\sum_{r=1}^s\frac{1}{n}\underset{m\rightarrow\infty}{\lim}~\log w_{rn}(\bd_{G^*_r})-sp\frac{\beta^*}{n}$ with respect to $\bd$, subject to 
$\underset{m\rightarrow\infty}{\lim}~\frac{k_r(\bd)}{m}=p$; $r=1,\ldots,s$.
Now note that as $p\in (0,s^{-1})$, the number of parameters associated with $G^*_r$ increases to infinity as $m\rightarrow\infty$.
Consequently, $w_{rn}(\bd_{G^*_r})$ is the posterior probability of the intersection of increasing number of events as $m$ increases. Hence, 
$\underset{m\rightarrow\infty}{\lim}~w_{rn}(\bd_{G^*_r})=w_{rn}(\bd^{\infty}_{G^*_r})$, say, where $\bd^{\infty}_{G^*_r}$ denotes the decision associated with 
infinite number of hypotheses in $G^*_r$. 
Thus, $\underset{m\rightarrow\infty}{\lim}~\log w_{rn}(\bd_{G^*_r})=\log w_{rn}(\bd^{\infty}_{G^*_r})$, and we are concerned
with the maximization of $p\sum_{r=1}^s\frac{1}{n}\log w_{rn}(\bd^{\infty}_{G^*_r})-sp\frac{\beta^*}{n}$ with respect to $\bd^{\infty}$ subject to
$\underset{m\rightarrow\infty}{\lim}~\frac{k_r(\bd^{\infty})}{m}=p$; $r=1,\ldots,s$.
Here $\bd^{\infty}$ denotes the entire infinite-dimensional decision configuration.

Now for any given $m\geq 1$, let $\hat\bd$ be the maximizer of $\frac{1}{mn}\sum_{r=1}^sk_{r}(\bd)\left(\log w_{rn}(\bd_{G^*_r})-\beta^*\right)$. 
By Lemma \ref{lemma:discovery}, $\sum_{i=1}^m\hat d_{i}$ is decreasing in $\beta^*$. Hence, any attainable proprotion $\frac{\sum_{i=1}^m\hat d_{i}}{m}$ can be achieved
by decreasing $\beta^*$, for any $m\geq 1$. In other words, by decreasing $\beta^*$ adequately one can achieve $\underset{m\rightarrow\infty}{\lim}~\frac{\sum_{i=1}^m\hat d_{i}}{m}=
\underset{m\rightarrow\infty}{\lim}~\frac{\sum_{i=1}^m d^t_{i}}{m}=\underset{m\rightarrow\infty}{\lim}~\sum_{r=1}^s\frac{k_r(\bd^t)}{m}=sp~(<1)$. Simultaneously
one can achieve $\underset{m\rightarrow\infty}{\lim}~\frac{\sum_{i\in G^*_r}\hat d_i}{m}=p$ by selecting that maximizer $\hat\bd$ such that $\sum_{i\in G^*_r}\hat d_i\approx mp$, for each $m\geq 1$.
Let us denote the corresponding $\beta^*$ by $\hat\beta^*$.

Now, by Shalizi's result, $\frac{1}{n}\log w_{rn}(\bd^{\infty}_{G^*_r})\approx -J\left(\bTheta_{\bd^{\infty}_{G^*_r}}\right)$,  
for sufficiently large $n$, where $\bTheta_{\bd^{\infty}_{G^*_r}}$ is the parameter space associated with $\bd^{\infty}_{G^*_r}$ in the same way as $\bTheta_{\bd}$
is associated with $\bd$.
Letting
$f_{\btheta^t_r}$ and $f_{\btheta_r}$ denote the marginal densities of the data $\bx_r$ associated with decisions $\bd^t_{G^*_r}$ and $\bd^{\infty}_{G^*_r}$, respectively, 
for $r=1,\ldots,s$, we obtain:
\begin{align}
J\left(\bTheta_{\bd^{\infty}_{G^*_r}}\right)
&=\underset{\btheta_r\in \bTheta_{\bd^{\infty}_{G^*_r}}}{\inf}\int\log\frac{f_{\btheta^t_r}(\bx_r)}{f_{\btheta_r}(\bx_r)}f^t_{\btheta^t_r}(\bx_r)dx_r\notag\\
&=\underset{\btheta_r\in \bTheta_{\bd^{\infty}_{G^*_r}}}{\inf}\int\log\frac{f_{\btheta^t_r}(\bx_r)}{f_{\btheta_r}(\bx_r)}\prod_{\ell=1}^s\left\{f^t_{\btheta^t_\ell}(\bx_\ell)dx_\ell\right\},\notag
\end{align}
so that, using disjointness of $\Theta^*_r$; $r=1,\ldots,s$, we obtain
\begin{align}
\sum_{r=1}^sJ\left(\bTheta_{\bd^{\infty}_{G^*_r}}\right)&=\sum_{r=1}^s\underset{\btheta_r\in \bTheta_{\bd^{\infty}_{G^*_r}}}{\inf}
\int\log\frac{f_{\btheta^t_r}(\bx_r)}{f_{\btheta_r}(\bx_r)}\prod_{\ell=1}^s\left\{f^t_{\btheta^t_\ell}(\bx_\ell)dx_\ell\right\}\notag\\
&=\underset{\btheta\in \bTheta_{\bd^{\infty}}}{\inf}\sum_{r=1}^s\int\log\frac{f_{\btheta^t_r}(\bx_r)}{f_{\btheta_r}(\bx_r)}\prod_{\ell=1}^s\left\{f^t_{\btheta^t_\ell}(\bx_\ell)dx_\ell\right\}\notag\\
&=\underset{\btheta\in\bTheta_{\bd^{\infty}}}{\inf}\int\log\frac{\prod_{r=1}^sf_{\btheta^t_r}(\bx_r)}{\prod_{r=1}^sf_{\btheta_r}(\bx_r)}\prod_{\ell=1}^s\left\{f^t_{\btheta^t_\ell}(\bx_\ell)dx_\ell\right\}\notag\\
&=J\left(\bTheta_{\bd^{\infty}}\right).
\label{eq:kl_J2}
\end{align}
It follows from (\ref{eq:kl_J2}) that $\frac{1}{n}\sum_{r=1}^s\log w_{rn}(\bd_{G^*_r})\approx -J\left(\bTheta_{\bd^{\infty}}\right)$ for large enough $n$.
This, along with the argument of the existence of an appropriate $\hat\beta^*$ in the previous paragraph shows that there exists $\hat\beta^*$ for which
our non-marginal method minimizes the essential infimum of the KL-divergence $J\left(\bTheta_{\bd^{\infty}}\right)$ from the 
true decision configuration among all decisions $\bd^{\infty}$ associated with any other multiple testing method satisfying \eqref{eq:constraint1}.

\subsection{Proof of Lemma \ref{lemma:lemmastationary}	}
Let $q(\dec^\ast,\dec)$ and $\alpha_i(\dec^\ast,\dec)$ be the probabilities that the decision configuration 
$\dec$ is proposed from $\dec^\ast$ and it is selected as the current configuration at step $i$ respectively.

Note that,
\begin{align} 
\pi_i(\dec)P_i(\dec^\ast|\dec)&= \pi_i(\dec)q(\dec^\ast,\dec)\alpha_i(\dec^\ast,\dec)\\
&= q(\dec^\ast,\dec)\min\{\pi_i(\dec),\pi_i(\dec^\ast)\}.			
\end{align}
Now, $q(\dec^\ast,\dec)$ is the probability that $\dec$ is proposed while the current decision configuration is $\dec^\ast$. 
Clearly, if an operation leads from $\dec$ to $\dec^\ast$ using our prescription, then we can revert back to $\dec$ 
using the same operation on $\dec^\ast$. So, for our proposal, $q(\dec^\ast,\dec)=q(\dec,\dec^\ast)$. Hence,
\begin{align}
\pi_i(\dec)P_i(\dec^\ast|\dec)&= q(\dec^\ast,\dec)\min\{\pi_i(\dec),\pi_i(\dec^\ast)\}\\
&=q(\dec,\dec^\ast)\min\{\pi_i(\dec),\pi_i(\dec^\ast)\}\\
&=\pi_i(\dec^\ast)P_i(\dec|\dec^\ast).
\end{align}
Thus, the Markov chain $P_i$ is \textit{reversible} with respect to $\pi_i$. Therefore,
\begin{equation}
\sum_{\dec\in\mathbb{D}} \pi_i(\dec)P_i(\dec^\ast|\dec)=\sum_{\dec\in\mathbb{D}} 
\pi_i(\dec^\ast)P_i(\dec|\dec^\ast)=\pi_i(\dec^\ast)\sum_{\dec\in\mathbb{D}} P_i(\dec|\dec^\ast)=\pi_i(\dec^\ast),
\end{equation}
showing that $P_i$ has stationary distribution $\pi_i$.

Note that, the Markov chain is also \textit{irreducible and aperiodic}. It is irreducible because as per our construction, 
there is always a positive probability to reach any state from any other state through only a finite number of alteration(s). 
The Markov chain is also aperiodic since the same state can be retained with positive probability at each step.
Moreover, because of finiteness of the state space, the above Markov chain is uniformly ergodic.
Hence, it follows that the algorithm converges in probability to the set of global maxima (see, for example, 
\ctn{rigorousanneal} and the references therein). 
\renewcommand\baselinestretch{1.3}
\normalsize
\bibliographystyle{natbib}
\bibliography{irmcmc}

\end{document}